# Homozygous *GRN* mutations: unexpected phenotypes and new insights into pathological and molecular mechanisms

**Authors:**

Vincent Huin, MD, PhD[a], Mathieu Barbier, PhD[a,b], Armand Bottani, MD[c], Johannes Alexander Lobrinus, MD[d], Fabienne Clot, PhD[e], Foudil Lamari, PharmD, PhD[f], Laureen Chat[e], Benoît Rucheton, PharmD[f], Frédérique Fluchère, MD[g], Stéphane Auvin, MD, PhD[h], Peter Myers, MD[i], Antoinette Gelot, MD, PhD[j], Agnès Camuzat, BSc[a], Catherine Caillaud, MD, PhD[k], Ludmila Jornéa, BSc[a], Sylvie Forlani, PhD[a], Dario Saracino, MD[a], Charles Duyckaerts, MD, PhD[l] Alexis Brice, MD[a,m] , Alexandra Durr, MD, PhD[a,m*], Isabelle Le Ber, MD, PhD[a,n]

[1] Sorbonne Université, Institut du Cerveau et de la Moelle épinière (ICM), AP-HP, INSERM, CNRS, University Hospital Pitié - Salpêtrière, Paris, France

[2] Department of Genetic Medicine, University Hospital of Geneva, Geneva, Switzerland

[3] Neuropathology Unit, University Hospital of Geneva, Geneva, Switzerland

[4] Department of Molecular and Cellular Neurogenetics, AP-HP, Pitié - Salpêtrière - Charles Foix University Hospitals, Paris, France

[5] AP-HP, Metabolic Biochemistry Unit, Department of Biochemistry of Neurometabolic Diseases, Pitié - Salpêtrière University Hospital, Paris, France

[6] AP-HM, Department of Neurology and Movement Disorders, La Timone, Clinical Neuroscience Unit, Aix-Marseille University, France

[7] AP-HP Department of Pediatric Neurology, Robert Debré University Hospital, F-75019, Paris, France

[8] Medical Office, Geneva, Switzerland

[9] Neuropathology, Department of Pathology, Trousseau Hospital, AP-HP, Paris, France and INMED INSERM U901 Luminy, Campus, Aix-Marseille University, France

[10] Biochemical, Metabolomical and Proteonomical Department, Necker-Enfants Malades University Hospital, AP-HP, F-75015, Paris, France

[11] Department of Neuropathology 'Escourolle', AP-HP, Pitié - Salpêtrière University Hospital, Paris, France

[12] AP-HP, National Reference Center for Rare Diseases 'Neurogenetics', Pitié - Salpêtrière University Hospital, Paris, France

[13] AP-HP, National Reference center 'rare and young dementias', IM2A, Pitié - Salpêtrière University Hospital, Paris, France

[*] **Correspondence:** Pr. Alexandra Durr, Institut du Cerveau et la Moelle épinière (ICM), Boulevard de l'hôpital, CS21414, 75646 PARIS CEDEX

alexandra.durr@icm-institute.org

Telephone: +33 (0)1 42 16 37 98

Fax: +33 (0)1 57 27 47 95

**Running title:**

New insights in homozygous *GRN* mutations

**ORCID number:**

Vincent Huin = 0000-0001-8201-5406

Mathieu Barbier = 0000-0002-5154-2163

Armand Bottani = 0000-0001-7146-1291

Foudil Lamari = 0000-0002-5104-2935

Benoit Rucheton = 0000-0002-5893-6595

Stéphane Auvin = 0000-0003-3874-9749

Antoinette Gelot = 0000-0002-3032-014X

Catherine Caillaud = 0000-0003-2712-7661

Dario Saracino = 0000-0002-4299-9743

Charles Duyckaerts = 0000-0002-2662-4810

Alexis Brice = 0000-0002-0941-3990

Alexandra Durr = 0000-0002-8921-7104

Isabelle Le Ber = 0000-0002-2508-5181

**Total word count of the manuscript excluding the title page, abstract, tables, acknowledgements and contributions, and references:** 6387

**Character count for the title:** 100

**Character count for the running title:** 40

**Number of references:** 52

**Number of tables:** 1

**Number of figures:** 7

**Number of supplemental tables:** 1

**Number of supplemental figures:** 3

## ABSTRACT

Homozygous mutations in the progranulin gene (*GRN*) are associated with neuronal ceroid lipofuscinosis 11 (CLN11), a rare lysosomal-storage disorder characterized by cerebellar ataxia, seizures, retinitis pigmentosa, and cognitive disorders, usually beginning between 13 and 25 years of age. This is a rare condition, previously reported in only four families. In contrast, heterozygous *GRN* mutations are a major cause of frontotemporal dementia associated with neuronal cytoplasmic TDP-43 inclusions. We identified homozygous *GRN* mutations in six new patients. The phenotypic spectrum is much broader than previously reported, with two remarkably distinct presentations, depending on the age of onset. A childhood/juvenile form is characterized by classical CLN11 symptoms at an early age at onset. Unexpectedly, other homozygous patients presented a distinct delayed phenotype of frontotemporal dementia and parkinsonism after 50 years; none had epilepsy or cerebellar ataxia. Another major finding of this study is that all *GRN* mutations may not have the same impact on progranulin protein synthesis. A hypomorphic effect of some mutations is supported by the presence of residual levels of plasma progranulin and low levels of normal transcript detected in one case with a homozygous splice-site mutation and late onset frontotemporal dementia. This is a new critical finding that must be considered in therapeutic trials based on replacement strategies. The first neuropathological study in a homozygous carrier provides new insights into the pathological mechanisms of the disease. Hallmarks of neuronal ceroid lipofuscinosis were present. The absence of TDP-43 cytoplasmic inclusions markedly differs from observations of heterozygous mutations, suggesting a pathological shift between lysosomal and TDP-43 pathologies depending on the mono or bi-allelic status. An intriguing observation was the loss of normal TDP-43 staining in the nucleus of some neurons, which could be the first stage of the TDP-43 pathological process preceding the formation of typical cytoplasmic inclusions. Finally, this study has important implications for genetic counselling and molecular diagnosis. Semi-dominant inheritance of *GRN* mutations implies that specific genetic counseling should be delivered to children and parents of CLN11 patients, as they are heterozygous carriers with a high risk of developing dementia. More broadly, this study illustrates the fact that genetic variants can lead to different phenotypes according to their mono- or bi-allelic state, which is a challenge for genetic diagnosis.

**Abbreviations: (**bv)FTD: behavioural variant FTD; FTD: frontotemporal dementia; GAPDH: glyceraldehyde 3-phosphate dehydrogenase; GRN: granulin precursor; IHC: immunohistochemistry; IR: immunoreactivity; LCL: lymphoblast cell lines; MLPA: Multiplex Ligation-dependent Probe Amplification; nfvPPA: nonfluent variant of primary progressive aphasia; NMD: non-mediated decay; PGRN: progranulin; svPPA: semantic variant of primary progressive aphasia; TDP-43: TAR DNA-binding protein 43; TMEM106B: Transmembrane Protein 106B.

## INTRODUCTION

Frontotemporal dementias (FTD) are rare degenerative dementias characterized by predominant behavioural changes (behavioural variant of FTD, bvFTD) or language disorders (semantic and nonfluent variants of primary progressive aphasia, svPPA and nfvPPA) usually occurring between 50 and 65 years (Rascovsky *et al.*, 2011; Gorno-Tempini *et al.*, 2011). FTD are associated with predominant frontal and temporal atrophy. Three pathological subtypes are defined according to the proteins (Tau, TDP-43 or FUS/FET) aggregated into neuronal inclusions (Mackenzie *et al.*, 2011). The most frequent neuropathological hallmarks are neuronal inclusions of TDP43, a nuclear protein which is translocated and aggregates in the cytoplasm in FTD patients.

A third of patients with FTD have a family history of the disease with an autosomal dominant mode of transmission. Most of them carry heterozygous pathogenic variants in *C9orf72*, *GRN* and *MAPT* genes. Among them, *GRN* (granulin precursor, OMIM *138945) is responsible for 20% of familial FTD (Baker *et al.*, 2006; Cruts *et al.*, 2006). *GRN* mutations lead to bvFTD or PPA phenotypes and, less frequently, to a corticobasal syndrome (CBS). Age at onset in heterozygous *GRN* carriers ranges from 40 to 85 years (Rademakers *et al.*, 2007; Le Ber *et al.*, 2008). Gestural apraxia, parkinsonism and visual hallucinations are present in 25 to 40% of the patients (Le Ber *et al.*, 2008). Most *GRN* pathogenic variants are null mutations leading to a partial loss of function caused by progranulin (PGRN) haploinsufficiency resulting from degradation of mutant RNA by non-mediated decay (NMD). Progranulin is a secreted protein and thus, decreased or undetectable plasma progranulin levels are highly predictive of pathogenic heterozygous GRN mutations. Different thresholds of 110 (Finch *et al.*, 2009), 74 (Ghidoni *et al.*, 2008), or more recently 61 mg/l (Ghidoni *et al.*, 2012; Galimberti *et al.*, 2018) have been established depending on the study. All heterozygous *GRN* mutations are associated with neuronal cytoplasmic and highly evocative ‘cat eyes’ nuclear TDP-43-positive inclusions.

Neuronal ceroid lipofuscinoses form a heterogeneous group of inherited lysosomal-storage disorders caused by at least 13 different genes (Mole *et al.*, 2019; Mukherjee *et al.*, 2019). They are characterized by abnormal accumulation of autofluorescent lipopigment in lysosomes detectable in various tissues, including the skin, retina, and brain. Neuronal ceroid lipofuscinosis type 11 (CLN11) is caused by homozygous *GRN* mutations, which lead to the complete loss of function of the protein. As expected, this genetic form is rare and has only been identified in five patients from four unrelated families (Smith *et al.*, 2012; Canafoglia *et al.*, 2014; Almeida *et al.*, 2016; Faber *et al.*, 2017; Kamate *et al.*, 2019). The cardinal features of these patients are cerebellar ataxia with cerebellar atrophy on brain MRI, retinitis pigmentosa, leading to visual loss, epilepsy, and progressive cognitive decline, with the age of onset from 13 to 25 years. Plasma PGRN is undetectable in patients with the CLN11 phenotype carrying homozygous *GRN* mutation (Smith *et al.*, 2012; Canafoglia *et al.*, 2014; Almeida *et al.*, 2016).

We report six patients from four unrelated families carrying homozygous *GRN* mutations with divergent and unexpected phenotypes, including bvFTD, parkinsonism, and variable ages of onset. We also report the first neuropathological findings associated with homozygous mutations. This study extends the mutation, phenotype, and neuropathological spectra of *GRN*-related disorders and suggests specific molecular and pathological mechanisms depending on the mono or bi-allelic genotype.

## PATIENTS AND METHODS

### Identification of the patients

Index cases were identified either as part of i) a diagnostic work-up for familial FTD, ii) targeted next-generation sequencing of all neuronal ceroid lipofuscinoses genes as part of the diagnostic work-up, or iii) exome sequencing in a research setting of the Spastic Paraplegia and Ataxia (SPATAX) network (https://spatax.wordpress.com/). Written informed consent was obtained from all subjects in accordance with French and Swiss ethics regulations. This study was approved by local French regulations [Paris Necker Ethics Committee approval (RBM 02-59 and RBM 03-48) to I.L.B. and A.D.].

### DNA molecular analyses

Molecular screening of *GRN* (NM_002087.2) was performed in the four index cases and in one affected relative by exome sequencing (Family AAR-427), targeted next-generation sequencing (Family NCL-001), or Sanger sequencing (FTD-1042 and FTDP-N12/1611). Phasing to confirm homozygosity of the variants was possible for two families. Multiplex ligation-dependent probe amplification (MLPA) was used to confirm the absence of a concomitant exon or gene deletion for the other two families. The single nucleotide polymorphism rs1990622 of the *TMEM106B* gene, of which the minor G-allele is associated with reduced penetrance in FTD caused by *GRN* mutations, was also genotyped in homozygous mutation carriers when possible, as previously described (Lattante *et al.*, 2014). Repeat expansion in the *C9orf72* gene was also ruled out for the two probands with the late-onset phenotype using a fluorescent repeat-primed PCR assay, as previously described (Le Ber *et al.*, 2013).

**The effect of the c.709-3C>G mutation on RNA splicing in family FTD-1042**

The c.709-3C>G mutation detected in one family (FTD-1042) is located in the 3' acceptor splice site of intron 7. We studied its biological effect on splicing and RNA levels. RNA was extracted from lymphoblast cell lines (LCL) of the homozygous proband (FTD-1042-1) and his two heterozygous children (FTD-1042-2, FTD-1042-3) using the RNeasy plus mini kit (Qiagen, Düsseldorf, Germany). LCL were treated or not with emetine, an inhibitor of NMD. Reverse transcription (RT) was performed using the SuperScript III first strand® kit (Life Technologies SAS, Villebon-sur-Yvette, France) and RT-PCR was performed on cDNA with the forward primer in exon 7 (5'-ATGGTTCTACCTGCTGTGAGCT-3') and the reverse primer in exon 8 (5'-GCAGGCAGCTTAGTGAGGAGGT-3'). The amplified fragments were sequenced on an ABI 3730 automated sequencer using the Big Dye 3.1 cycle sequencing kit (Applied Biosystems, Foster City, CA). The sequencing data were analyzed using SeqScape 3 software (Applied Biosystems, Foster City, CA).

Quantitative real-time RT-PCR (qRT-PCR) was performed on the cDNA of the proband and his two children. *GRN* cDNA levels were determined using the FastStart Essential DNA Green Master kit (Roche, Indianapolis, IN) and a LightCycler® 480 (Roche, Indianapolis, IN) using the forward primer (5'-CAACGCCACCTGCTGC-3'), located at the junction between exons 7 and 8, and the reverse primer (5'-TCCGTGGTAGCGTTCTCC-3') in exon 8. Samples were tested in triplicate and the signal normalized against that of glyceraldehyde 3-phosphate dehydrogenase mRNA ($n = 4$). The relative transcription levels of *GRN* mRNA were calculated using the $2^{-\Delta\Delta Ct}$ method. As controls, we used RNA extracted from LCL of two healthy individuals not carrying *GRN* mutations and an unrelated FTD patient carrying the heterozygous c.813_816del, p.(Thr272Serfs*10) mutation in exon 8.

**Plasma progranulin assay**

Plasma progranulin levels were measured by ELISA using the progranulin-human-ELISA kit (Adipogen, Coger SAS, France), according to the manufacturer's instructions. The antibody epitopes are located at the C-terminus of progranulin, after the Granulin E/7 sequence. The analytical performances of the kit have been previously studied in our laboratory, as the normal values vary depending on the different studies and laboratories. A cut-off of >72 µg/L defining the normal values was established in our laboratory, and considered in this study. More information is provided in supplementary methods.

**Neuropathology**

Neuropathological examination of the proband (patient 1) from family AAR-427 was performed. The brain was formalin-fixed for four weeks before being examined using standard methods. Multiple samples were paraffin embedded (frontal, temporal, parietal and occipital cortex, hippocampus, cingular gyrus, caudate nucleus, putamen, pallidum, thalamus, cerebellar hemisphere, vermis, and dentate nucleus). Thin sections (2 and 5 µm thick) were mounted on slides and stained with hematoxylin and eosin, luxol fast blue for myelin, and periodic acid Schiff. Immunohistochemistry (IHC) was performed with various antibodies against TDP-43, phosphorylated TDP-43 (pTDP-43), Aβ, phosphorylated tau, ubiquitin, neurofilament and alpha-synuclein, and myelin basic protein (see supplementary table S1), as described (Seilhean *et al.*, 2011). Selected areas were fixed in glutaraldehyde and embedded in Epon for electron microscopy examination.

**Data availability**

The data that support the findings of this study are available from the corresponding author, upon reasonable request.

## RESULTS

## Clinical description

The phenotypic characteristics and genotypes of the six patients, as well as of patients of the four families with bi-allelic *GRN* variants previously reported, are summarized in table 1. Family trees are shown in figure 1.

### *Late-onset phenotypes*

**FTD-1042.** A 56-year-old woman initially had visual hallucinations (animals). Two years later, she developed language and behavioural disorders. She exhibited compulsive shopping, stereotyped behaviours, aggressiveness, and hyperphagia. At age 58, she had severe motor stereotypies, prehension and imitation behaviours, and grasping reflex associated with executive dysfunction. She rapidly presented major comprehension disorders and mutism. A diagnosis of bvFTD was made based on the international diagnostic criteria (Rascovsky *et al.*, 2011). Examination revealed an akinetic-rigid syndrome, pyramidal spasticity of the four limbs, and right extensor plantar responses. Brain MRI showed cortico-subcortical atrophy, predominating in the right temporal lobe. The behavioural disorders and brain MRI were consistent with a 'probable' bvFTD according to international clinical criteria (Rascovsky *et al.*, 2011). Two independent plasma PGRN assays showed low values of 30 and 39 µg/L.

Her mother died at the age of 70. No neurological disorder was reported in her mother's family. Familial information was censored for the proband's father. Sanger sequencing revealed that the proband carried a homozygous splice pathogenic variant, c.709-3C>G, p.?, in intron 7 of *GRN*. There were no *GRN* whole gene or exon deletions detected by MLPA and no mutation in *C9orf72*. The proband had the *TMEM106B* rs1990622 AA genotype. Two children (FTD-1042-2, FTD-1042-3), asymptomatic at the ages of 22 and 27, carried heterozygous *GRN* mutations and had low plasma PGRN levels (53 and 56 µg/L, respectively).

**FTDP-N12/1611.** A 61-year-old man (patient one) presented with atypical parkinsonian syndrome and language and cognitive deterioration. He had suffered from retinitis pigmentosa since the age of 44 and was almost blind at the time of examination. Parkinsonian syndrome was manifested by tremor in the upper limbs, postural instability, and gait impairment with rapid shuffling steps. Examination at age 63 revealed a severe akineto-rigid syndrome with bilateral postural tremor (predominating in the left upper limb) and axial signs (hypophonia, dysarthria, freezing, and abnormal retropulsion test). There was no significant response to L-Dopa treatment, as the Unified Parkinson's Disease Rating Scale (UPDRS) score III was 56/108 and 48/108 before and during the acute L-DOPA test. At the same time, he presented cognitive and behavioural disorders characterized by severe apathy and social withdrawal with diminished interest to others. He progressively developed loss of mental flexibility, marked irritability and aggressiveness, proffering coarse words and insults to his family members. He showed behavioural disinhibition with marked impulsivity, inappropriate social conducts, as well as hyperorality with binge eating and bulimia, frequently rushing on food and drinks. He also presented puerilism and marked emotional lability with frequent inappropriate crying for innocuous remarks or trivial events. Neuropsychological examination at age 63 showed a large decrease in global cognitive efficiency and frontal cognitive dysfunction. The Mini-Mental State examination score was 19/25 and the Mattis Dementia Rating Scale score was 58/90 (visual items were impossible to test). The patient showed deficits in executive functions, including deficits in attention, working memory, planning and conceptualization, and reduced verbal fluency. CT-scan revealed bilateral frontal, left perisylvian, and parietal atrophy (Figure 2A). 18F-Fluorodeoxyglucose positron emission tomography (FDG-PET) showed bilateral hypometabolism in the anterior cingular cortex, medial prefrontal cortex, extending to the parieto-occipital regions (Figure 2B-D). The cerebellum was normal. The behavioural disorders (disinhibition, apathy, hyperorality and loss of empathy), associated with cognitive decline marked by a predominant executive deficit and frontal involvement on neuroimaging, were consistent with 'probable' bvFTD, based on international diagnostic criteria (Rascovsky *et al.*, 2011), associated with parkinsonism. At age 65, the patient had visual hallucinations. Plasma PGRN was undetectable. He progressively loss his autonomy, became mute age 66, and died at age 68.

One sib (patient 2) had visual deficit at age 30. A diagnosis of retinitis pigmentosa was done at age 33 and he became blind at 36 of years. He developed behavioural changes at age 58. The family initially noticed inappropriate joviality when he had learnt his father's decease. At age 59, behavioural disorders were characterized by altered social conducts with disinhibition, joviality, unmotivated laughs, and loss of empathy. Weird behaviours were present: for example, he was always moving along the walls and repeatedly moved furniture at home, inappropriately. He also developed apathy that rapidly became severe, impulsivity and had eating changes associating binge eating, bulimia and oral exploration of inedible items such as pencils, ties or soaps. He had fixed ideas, perseverative and stereotyped sentences. Attentional and judgement disorders were present. The patient refused to undergo formal neuropsychological evaluations. Brain MRI revealed frontal and temporal atrophy. He had no cerebellar ataxia, no parkinsonian symptoms, no movement disorders, no epileptic seizures or other neurological disorders. A bvFTD was diagnosed at age 60, based on marked behavioural changes (behavioural disinhibition, hyperorality, apathy, stereotypies of speech) as well as brain MRI that were consistent with criteria of 'probable bvFTD' (Rascovsky *et al.*, 2011). Strinkingly, despite being blind, he possibly had visual hallucinations as he frequently tried to catch invisible things in the air. He progressively loss his autonomy, became mute at age 63, and died at age 64. No genetic testing could be performed.

Another sib (patient 4) presented visual deficit at age 18. A diagnosis of retinitis pigmentosa was done at age 30. He was blind at age 50. He had no neurological signs until age 59 (41 years after onset of visual symptoms). He developed frontal behavioural and cognitive disorders with tremor and akineto-rigid syndrome. Behavioural disorders were characterized by important apathy. He didn't initiate any action at home, staying sited at the same place without eating during several days in the absence of his spouse. Personal care was altered. He presented weird behaviours and repeatedly broke the glass of the window to go out his bedroom. He also presented affective and emotional blunting, and marked loss of empathy. For example, he showed complete indifference to a child who announced he suffered a cancer. He has mild eating changes with more important consummation of candies. Attentional deficits and altered judgment in daily-life were consistent with a frontal cognitive syndrome, but no formal neuropsychological testing could be performed. He did not completely fulfill the diagnostic criteria for bvFTD but the association of apathy, loss of empathy and some eating changes were strongly evocative of a frontal type of dementia. Analysis of Alzheimer's disease biomarkers in cerebrospinal fluid were not consistent with an amyloid pathology. He had no cerebellar ataxia, epileptic seizures or other neurological disorders. He progressively became mute, loss his autonomy, developed swallowing disorders and had a gastrostomy at age 64.

None of the three affected sibs of this family had cerebellar ataxia or epilepsy. Another sib (subject 3) was at age 65. Proband's father had developed progressive dementia at age 60 and died at age 66. His mother had a clinical diagnosis of Parkinson's disease at age 70 and died at age 83, and two maternal aunts had dementia at age 65. A consanguinity was possible as their families originated from the same small geographic region.

Sanger sequencing showed the index case (patient one) and a sib (patient three) a to carry a homozygous pathogenic variant, c.443_444del p.(Gly148Valfs*11), in exon 5 of *GRN*. There were no *GRN* whole gene or exon deletions identified by MLPA and no mutation in the *C9orf72* gene.

***Early-onset phenotypes***

**AAR-427.** Patient 1. This right-handed woman presented generalized tonic-clonic seizures at age 16, with a poor response to antiepileptic drugs. EEG showed diffuse dysrhythmia with anterior predominance and rare spike-wave discharges that were predominant in the left frontal region. At age 19, she developed bilateral visual loss caused by retinitis pigmentosa with cystoid macular edema (bilateral visual acuity 2/10) and bilateral cataracts. One year later, she presented irritability, impulsiveness, and a deficit in mental flexibility. Cerebellar gait disorder and dysarthria progressively developed at age 22. Brain MRI at age 22 showed cerebellar atrophy and mild diffuse bilateral cortical atrophy (Figure 2E-F). Neuropsychological examination at age 23 showed moderate loss of manners with deficits in frontal executive functions (mental flexibility, inhibition, and initiation), working memory, and reduced verbal fluency. Dyscalculia and constructive apraxia were present. At age 25, she presented visual hallucinations (various animals, road signs) not related to seizures. The cerebellar ataxia worsened, with saccadic pursuit, nystagmus, severe dysarthria, and dysphagia. She had generalized myoclonus at age 26. Cerebral CT scan at age 27 showed progression of cerebellar atrophy, with only discrete bilateral frontal sulcus enlargement (Figure 2G-H). She died at age 27 from food inhalation during an epileptic seizure. The clinical diagnosis was evocative of neuronal ceroid lipofuscinosis. An autopsy was performed.

Patient 2. Her sister went to school until age 18. She had generalized tonic-clonic seizures at age 12, which were treated with sodium valproate, and developed discrete and insidious cerebellar gait disorders. Cerebellar ataxia was evident at age 19. Progressive visual loss developed at age 22, caused by bilateral retinitis pigmentosa, macular edema, and cataracts. Cerebellar ataxia rapidly worsened by age 24, after childbirth. Walking difficulties, dysarthria, and saccadic pursuit were associated with cognitive deficit and psychomotor slowness. Neuropsychological examination at age 27 showed deficits in executive functions and working and episodic memory and word-finding difficulties. She had myoclonus at age 26 and visual hallucinations at age 28. At age 36, the patient was blind, presented severe dysarthria and dysphagia, and needed a wheelchair. Brain MRI at ages 22 and 29 and CT scan at age 34 showed progressive cerebellar atrophy and mild cortical atrophy, predominating in the bilateral frontal and right perisylvian regions (Figure 2I-L). Vacuolar inclusions were observed in lymphocytes from peripheral blood in both sisters (Figure 3A).

Their parents, aged 57 and 61, were asymptomatic. There was possible but undocumented consanguinity, as they originated from the same small town in Portugal. Plasma PGRN levels were 68 and 53 μg/L in the parents, respectively, and it was undetectable in the proband's sister. Exome and Sanger sequencing showed the two sisters to carry a homozygous frameshift pathogenic variant, c.768_769dup, p.(Gln257Profs*27), in exon 8 of *GRN*, inherited from their two heterozygous parents. The proband had the *TMEM106B* rs1990622 GG genotype, while her sister and their parents all had the AG genotype.

**NCL-001.** The proband presented with generalized tonic-clonic seizures at the age of seven, rapidly associated with cerebellar ataxia and memory and attention deficits, leading to learning disabilities. She had two seizures per year. Electroencephalography showed a photoparoxysmal response at low frequency (Supplemental Figure S1). She was successively treated with sodium valproate, levetiracetam, and topiramate, with limited effectiveness against the frequency of seizures. Brain MRI at age eight showed cerebellar atrophy predominating in the cerebellar hemispheres (Figure 2M-N). The learning disabilities worsened with disease progression, leading to severe school difficulties. Brain MRI at age 12 showed progression of cerebellar atrophy and moderate bilateral frontal atrophy, with dilated ventricles (Figure 2O-P). A diagnosis of progressive epileptic encephalopathy was made. At the age of 13, she complained of visual loss. Visual evoked potentials were decreased. Electroretinogram evaluation and fundoscopy were normal. A skin biopsy at age 14 showed rare vesicles surrounded by a membrane, of which the content was electron dense, in the cytoplasm of endothelial cells (Figure 3B). These vesicles were filled with membrane formations which were morphology evocative of curvilinear profiles, although they were more compact and formed a nearly complete circle (Figure 3C), instead of the classical aspect of curvilinear profiles which are uniformly curved, short, thin lamellar stacks of alternative dark and light lines with only minor variation in the degree of bending (Mole *et al.*, 2011; Anderson *et al.*, 2013). Although not completely classical, these findings led to the genetic testing of the neuronal ceroid lipofuscinoses genes.

Her parents originated from Pakistan and were first cousins. They were asymptomatic at age 46 and 47. Molecular screening of the neuronal ceroid lipofuscinoses genes showed the proband to carry a homozygous pathogenic variant, c.1A>T, p.(Met1?), in exon 2 of *GRN*. Plasma PGRN in the proband was undetectable.

**Effect of the *GRN* mutation c.709-3C>G on RNA splicing (family FTD-1042)**

The splice site c.709-3C>G mutation detected in the homozygous state in one family (FTD-1042) was located in the 3' acceptor splice site of intron 7 (Supplementary Figure S2A). This mutation was predicted to alter the acceptor site by GeneSplicer, NNSPLICE, MaxEntScan, and SpliceSiteFinder-like softwares and to weaken the intron 7 acceptor splice site by HSF software (strength of the acceptor site with the wild-type sequence = 89.13 and that with the mutation = 78.83, supplementary Figure S2B).

We studied the biological effect of the c.709-3C>G mutation on splicing and RNA expression levels in LCL. A single fragment, corresponding to the expected size of 178 bp, was present in non-mutated controls. We detected two fragments from LCL treated with emetine in the patient carrying the homozygous c.709-3C>G mutation: one of the expected size of 178 bp and another larger fragment of 414 bp (Figure 4A).

Sequencing of the amplified fragments showed retention of full-length intron 7 of 236 bp (Figure 4B) in the larger fragment. It confirmed that the intronic mutation c.709-3C>G altered splicing of exon 8, leading to a frameshift at codon Ala237 and a premature stop codon at position 98 downstream p.(Ala237Valfs*98). We did not detect the larger fragment in untreated cells, showing that the mutant mRNA was degraded by NMD, as expected.

Quantitative real-time RT-PCR showed markedly decreased levels of normal *GRN* transcript in the homozygous patient (11.2-fold) relative to a non-mutated control. Normal *GRN* transcript levels were decreased to a lesser degree in his heterozygous children (3.2 to 3.4-fold) relative to a non-mutated control (Figure 4C). The positive control carrying a heterozygous frameshift c.813_816del, p.(Thr272Serfs*10) mutation in exon 8 showed intermediate levels, higher than those of homozygous individuals but lower than heterozygous carriers of family FTD-1042.

**Neuropathological examination**

The total brain weight of patient one of family AAR-427 was 1,315 grams. Macroscopic examination revealed cerebellar atrophy involving both hemispheres and the vermis, without important atrophy of the supra-tentorial structures (Figure 5A). The medulla oblongata showed pallor of olivary nuclei (Figure 5B). Ultrastructural analysis of the temporal cortex neurons showed numerous lysosomal deposits with curvilinear-like profiles and very few small fingerprint profiles (Figure 5C-F).

Microscopic examination showed that all neurons in the CA4 sectors were ballooned; their nucleus was on the side and their cell body contained abundant, luxol-positive granules of lipofuscin in which formalin pigments were apparent (Figure 6A-B). The number of ballooned neurons was lower but still high in the CA1 sector. They were rare in the subiculum, in which most neurons were hypoxic. In the neocortex, there were a few ballooned neurons among hypoxic ones. The entire thickness of the neocortex was vacuolated – a possible consequence of hypoxia - but the neuronal loss was mild.

TDP-43 IHC showed the nuclei of the ballooned neurons to be devoid of immunoreactivity (Figure 6C-D). The cytoplasm of the cell bodies was diversely labeled: in some cases, the labeling was diffuse, whereas in others, it was concentrated around the nucleus (Figure 6D). The cells were negative for phosphorylated TDP-43. Some exceptional intracytoplasmic granules were immunopositive with ubiquitin antibodies. No inclusions similar to those associated with fronto-temporal lobar degeneration were found. Aβ, tau and α-synuclein IHC were negative.

In the cerebellum, the number of Purkinje cells was moderately decreased (Figure 6E). In contrast, neuronal loss was severe in the granular layer. There was exceptional grumose degeneration of the dentate nucleus; neuronal loss was nearly complete and there were numerous clusters of myelinated fibers (Figure 6F) arranged in spherical structures.

## DISCUSSION

We report four novel families with *GRN*-related disorders and bi-allelic pathogenic variants. This study has several implications for the clinical practice and understanding of pathological and molecular mechanisms of *GRN* diseases.

First, we confirmed that homozygous *GRN* mutations are involved in CLN11, as only four families have been reported until now (Smith *et al.*, 2012; Canafoglia *et al.*, 2014; Almeida *et al.*, 2016; Faber *et al.*, 2017; Kamate *et al.*, 2019). All mutations identified here have been previously detected in the heterozygous state in FTD patients (Benussi *et al.*, 2009; Almeida *et al.*, 2014; Perry *et al.*, 2013; Pottier *et al.*, 2018) and one, p.(Gln257Profs*27), in the homozygous state in another CLN11 patient (Faber *et al.*, 2017).

Second, this study sheds further light on the tight links between PGRN and lysosomes in neurons. How progranulin deficiency contributes to neurodegeneration is still unclear. Progranulin has multiple functions as a neurotrophic factor and in neuroinflammation, two important biological pathways that contribute to neurodegeneration in progranulin-deficient models (Van Damme *et al.*, 2008; Martens *et al.*, 2012). Progranulin also plays a major role in lysosomes, where it is cleaved into granulins (Hu *et al.*, 2010; Zhou *et al.*, 2017*a*) and modulates proteins (beta-glucocerebrosidase, cathepsin D) involved in lysosomal disorders (Jian *et al.*, 2016; Zhou *et al.*, 2017*b*). An overlap between FTD and CLN was previously shown in *Grn*-knockout mice, a model of *GRN*-related FTD that develops lysosomal vacuolization and lipofuscin accumulation (Ahmed *et al.*, 2010). More importantly, evidence that lysosomal dysfunction contributes to the pathological mechanisms of *GRN* came from the description of five CLN11 patients with homozygous mutations and symptoms of lysosomal storage disorders (Smith *et al.*, 2012; Canafoglia *et al.*, 2014; Almeida *et al.*, 2016; Faber *et al.*, 2017; Kamate *et al.*, 2019). However, whether lysosomal dysfunction in brain tissue directly contributes to neurological degenerative symptoms in these patients has not been documented until now. The hallmarks of lysosomal storage disorders in the brain tissue of our autopsied case provides the first evidence that complete PGRN deficiency causes lysosomal dysfunction in the brains of patients.

Third, we extend the phenotypic spectrum associated with bi-allelic *GRN* mutations, as many clinical features in our study are atypical of CLN11.

### *Homozygous GRN mutations lead to large phenotypic variability: from childhood to late-adulthood onset forms*

All reported CLN11 cases developed their first symptoms between the ages of 13 and 25 years. In our study, the age of onset was extremely heterogeneous, ranging from seven years to onset during the sixth decade. We describe two distinct clinical phenotypes according to the age of onset. The childhood/juvenile-onset form is characterized by generalized tonic-clonic epilepsy, cerebellar ataxia, and retinitis pigmentosa, which are later associated with frontal cognitive dysfunction. This form is consistent with that of the CLN11 patients described previously, except for the single case of onset at seven years, which is the youngest to date.

Unexpectedly, other homozygous patients described in this study presented later onset (FTD-1042, FTDP-N12/1611). They have developed a remarkably distinct and less severe neurological phenotype consistent with bvFTD ± parkinsonism occurring after the age of 50. Proband from family FTD-1042, and patients 1 and 2 from family FTDP-N12/1611, have developed behavioural and cognitive symptoms responding to the international clinical criteria of 'probable' bvFTD (Rascovsky *et al.*, 2011). The patient 4 from family FTDP-N12/1611 developed frontal dysfunction too, but not strictly fitting the criteria of 'possible' bvFTD. Parkinsonian symptoms of variable severity were present in three of them, since disease onset or during its progression.

The patient FTD-1042 had no other neurological or extra-neurological symptoms compatible with CLN11 classical phenotype. It is well known that heterozygous *GRN* mutations lead to FTD (Cruts *et al.*, 2006; Baker *et al.*, 2006) but this is the first evidence that homozygous mutations may be evocated in the case of a bvFTD phenotype as well. The three patients from family FTDP-N12/1611 developed retinitis pigmentosa that remained isolated a long time (17 to 41 years) before the occurrence of frontal dysfunction. Conversely to younger cases, none had developed other cardinal features of neuronal ceroid lipofuscinosis such as cerebellar ataxia, cerebellar atrophy or epilepsy. This association is also evocative of a protracted late form of neuronal ceroid lipofuscinosis, but pathological examination was not available in these cases to allow a definite diagnosis. A late-onset phenotype has been described previously for other forms of neuronal ceroid lipofuscinosis (Lauronen *et al.*, 1999; van Diggelen *et al.*, 2010; Xin *et al.*, 2010; Arsov *et al.*, 2011; Smith *et al.*, 2013). However, the features in patients from family FTDP-N12/1611 were remarkable by the unusually long intervals (from 17 to 41 years) between retinitis pigmentosa and the onset of dementia, and by the absence of any other cardinal features of neuronal ceroid lipofuscinosis, such as cerebellar ataxia, cerebellar atrophy, or epilepsy. Finally, these observations mirror that of two Belgian sibs with bvFTD related to adult-onset CLN13 caused by homozygous mutation in the *CTSF* gene (van der Zee *et al.*, 2016). This study thus extends the indications of genetic testing, going from childhood/juvenile ceroid lipofuscinosis to old adult-onset bvFTD with parkinsonism.

***Visual hallucinations are 'red-flags' in CLN and FTD patients***

Visual hallucinations were present in both childhood/juvenile and old-adulthood forms and are illustrative of a continuum between CLN11 and FTD. Complex visual hallucinations, mainly consisting of seeing people and animals, are frequent (25%) in heterozygous carriers (Le Ber *et al.*, 2008) but has not been described in CLN11 patients. Thus far, only two affected siblings were previously reported to have palinopsia and a hyperexcitable occipital cortex on electrophysiological studies (Canafoglia *et al.*, 2014). In our study, 4/6 patients had visual hallucinations. This high frequency in young patients with the CLN11 phenotype (2/3) suggests that they may represent "red flag" signs for homozygous *GRN* mutations, not only in patients with neuronal ceroid lipofuscinosis but also those with cerebellar ataxia at onset. We propose to include *GRN* in the genetic screening of autosomal recessive cerebellar ataxia, when associated with hallucinations. Although the pathophysiological mechanisms of visual hallucinations of *GRN* patients are not yet clear, they may be related to the lesions of retinitis pigmentosa in homozygous carriers and the less severe and infraclinical lesions of retinal lipofuscinosis that have been detected in heterozygous carriers (Ward *et al.*, 2017).

***Homozygous mutations are unexpected cause of late-onset FTD: the effect of hypomorphic mutations?***

Another major finding of this study is the evidence that all *GRN* mutations do not have the same impact on PGRN synthesis. Although most pathogenic mutations lead to haploinsufficiency, some may have a milder functional impact. Indeed, it was surprising that three homozygous carriers developed FTD at the same age as heterozygous carriers (40-85 years). Remarkably, one patient (FTD-1042-1) had plasma progranulin levels in the same range as heterozygous carriers (32 µg/l). As most pathogenic *GRN* mutations lead to the degradation of mutant RNA by non-sense mediated decay (Cruts *et al.*, 2006; Baker *et al.*, 2006), it would be expected that homozygous mutations lead to a complete loss of function without any residual PGRN protein. This was illustrated by undetectable plasma progranulin in CLN11 patients (Smith *et al.*, 2012; Canafoglia *et al.*, 2014; Almeida *et al.*, 2016).

However, the pathophysiological mechanism in *GRN*-related disorders is probably more complex, not only depending on PGRN deficit, as undetectable levels were also found in three heterozygous carriers (Calvi *et al.*, 2015; Galimberti *et al.*, 2016), and as one heterozygous carrier in this study had plasma PGRN level just under the limit of normal value (68 µg/L).

In patient one of family FTD-1042, we demonstrated that the causal c.709-3C>G splice mutation in homozygous state produces two different transcripts. The mutant transcript of 414 bp resulted from the retention of intron 7 and was degraded by NMD, as shown by emetine treatment. The occurrence of intron 7 retention, rather than exon 8 skipping, was probably linked to weakening of the intron 7 acceptor splice site, a short intron length (236 bp), and high intronic GC content (64.8% GC), which are three mechanisms known to favor intron retention (Sibley *et al.*, 2016).

Strikingly, we also detected a low level of normally spliced mRNA. This shows that the c.709-3C>G mutation does not completely inactivate the intron 7 acceptor splice site but produces a weaker, but still functional, acceptor site, as predicted by *in silico* software and expected for mutations that do not affect the canonical splice site. We demonstrated that the mutation, although in the homozygous state, allows the synthesis of a residual amount of normal *GRN* transcript and protein, consistent with low levels of normal mRNA and the plasma progranulin levels detected in the homozygous proband of family FTD-1042. Importantly, this implies that all *GRN* variants do not have a similar impact on PGRN synthesis. Some hypomorphic variants, such as the c.709-3C>G mutation, may lead to higher levels of PGRN synthesis, a milder phenotype, and an older age of onset than other conventional mutations. This is a new and potentially important finding to consider in the development of therapeutic trials based on replacement strategies.

However, this is probably not the only mechanism, as one patient with the same mutation in the heterozygous state developed FTD in the same age range as FTD-1042 (Benussi *et al.*, 2009). Furthermore, this does not appear to be relevant for all patients of this series, as plasma progranulin was undetectable in at least one other case with late onset FTD (FTDP-N12/1611). In this case, carrying the c.443_444del, p.(Gly148Valfs*11) mutation in exon 5 may lead to degradation of the mutant RNA by NMD. Alternatively, it could produce a partially functional residual truncated protein not detected by the ELISA Adipogen kit (epitopes are located in domain E, at the extremity of the C-terminus). LCL of this patient were not available to study the impact of the mutation in more detail.

Finally, mechanisms are probably more complex and additional modifiers, including genetic factors, may also influence the age of onset. The protective minor G-allele of rs1990622 in *TMEM106B* is associated with reduced disease penetrance in heterozygous *GRN* patients (Van Deerlin *et al.*, 2010; Finch *et al.*, 2011; Lattante *et al.*, 2014) and should be investigated in homozygous carriers as well. In our study, one of the late-onset patients (FTD-1042) carried the non-protective AA genotype, and two sisters had similar early-onset phenotypes (AAR-427) but different genotypes (GG and AG). Therefore, these results do not support rs1990622 contribution and suggest the involvement of other modifiers in *GRN*-related disorders.

***Mono vs bi-allelic GRN diseases: a possible phase shift between TDP-43 and lysosomal pathologies?***

We report the first neuropathological findings in a homozygous p.(Gln257Profs*27) *GRN* carrier with a juvenile CLN11 phenotype. In heterozygous *GRN* patients, neuropathological findings are characterized by cytoplasmic and intra-nuclear TDP-43-immunoreactive inclusions in neurons (Mackenzie *et al.*, 2011). The same inclusions were observed in one patient carrying the heterozygous p.(Gln257Profs*27) mutation (Pires *et al.*, 2013). In our case, hallmarks of neuronal ceroid lipofuscinoses including deposits with curvilinear profiles were observed. There was no major neuronal loss in the neocortex and no typical cytoplasmic TDP-43 inclusions. These major differences show there is a phase shift between lysosomal and TDP-43 pathologies in *GRN*-related disorders, depending on the presence of the mono or bi-allelic genotype.

The most intriguing finding was the loss of normal nuclear TDP-43 staining observed in some neurons. This was sometimes associated with abnormal immunoreactivity in the cytoplasm forming a ring around the nucleus. There is still a debate concerning whether TDP-43 induces neuronal loss by a toxic gain of function of cytoplasmic aggregates or by the loss of its normal function in the nucleus (Lee *et al.*, 2012). In this case, the absence of typical TDP-43 aggregates suggest that they may not be necessary to promote neuronal dysfunction in *GRN*-related disorders. Alternatively, the depletion of nuclear TDP-43 in some neurons could be a stigma of the presymptomatic stage of FTD in this young CLN11 patient, who died prematurely of a seizure. It mirrors the observation of the loss of nuclear TDP-43 that preceded the formation of cytoplasmic inclusions in a presymptomatic *C9orf72* carrier by several years (Vatsavayai *et al.*, 2016). Finally, it is possible that the ring deposits observed around the nucleus in some neurons form 'pre-inclusions', in which TDP-43 aggregation is initiated, before forming typical inclusions later as the disease progresses (Brandmeir *et al.*, 2008).

### *Impact on diagnosis and genetic counselling*

Our study will have practical consequences for the diagnosis of *GRN*-related disorders and the counseling of families. Bi-allelic *GRN* mutations upset inheritance models of FTD and genetic counseling. Most neuronal ceroid lipofuscinoses are autosomal recessive diseases, with clinical expression limited to bi-allelic variants carriers. Conversely, CLN11 is characterized by a semi-dominant transmission, the presence of heterozygous *GRN* mutations being sufficient to develop FTD with a penetrance of about 90% by the age of 75 years (Cruts *et al.*, 2006; Le Ber *et al.*, 2008). First degree relatives of bi-allelic *GRN* carriers thus have high risk of developing dementia later in life, as they are obligate carriers of heterozygous mutations. This implies that specific genetic counseling should be delivered to family members, not only to the children, but also to the parents of CLN11 patients.

The semi-dominant mode of transmission of *GRN*-related diseases parallels that of other neurogenetic and metabolic diseases. For example, mutations in *GRID2* and *SETX* in cerebellar ataxia and hereditary spastic paraplegia (HSP) have been linked to variable ages of onset (from congenital to adult-onset) and distinct clinical presentations (from pure to complex HSP for *GRID2* and from cerebellar ataxia to amyotrophic lateral sclerosis for *SETX*), depending on the mono or bi-allelic genotype (Coutelier *et al.*, 2015; Groh *et al.*, 2017). The growing evidence that there is no strict 'one gene, one disease' relationship and that the same variant in a given gene can produce very different phenotypes, depending on its mono- or bi-allelic state, is a challenge for the identification and interpretation of causative mutations in still unexplained genetic diseases. This study illustrates the utility of applying next generation sequencing to patients with an unclear genetic aetiology to detect infrequent occurrences, such as homozygosity in rare dominant diseases. It also emphasizes the limitations of genetic testing solely guided by inheritance patterns.

### *Conclusion*

In summary, this study breaks down the barrier between CLN11 and FTD, as homozygous GRN mutations can lead to both disorders. These two diseases are extreme phenotypes of a common spectrum of disorders caused by bi-allelic mutations (Figure 7).

Importantly, we show that some GRN mutations have a hypomorphic effect, a major finding to consider for future replacement therapies. However, this does not explain all the phenotypic variability and additional modifying factors probably also influence progranulin expression levels and the clinical outcome. The comparison of patients with CLN11 and those with FTD could provide important information about such modifiers. Further investigation of these patients may offer an original way to search for environmental or genetic factors that could explain different clinical outcomes toward neuronal ceroid lipofuscinosis or FTD or different ages of onset in these two neurodegenerative diseases.

## ACKNOWLEDGEMENTS

We are grateful to all the patients and family members for their participation in this study. We thank Dr C. Hillion (CHU Genève, Switzerland) for providing key clinical information on the earlier care of family AAR-427, and the DNA and cell bank of the ICM (ICM, Paris), Kathy Larcher (UF de Neurogénétique, Pitié-Salpêtrière hospital, Paris), Sandrine Noël (UF de Neurogénétique, Pitié-Salpêtrière hospital, Paris), Isabelle David (UF de Neurogénétique, Pitié-Salpêtrière hospital, Paris), Imen Benyounes (Laboratoire de biochimie métabolique, Pitié-Salpêtrière hospital, Paris) and Jean-Philippe Puech (Biochimie Métabolomique et Protéomique, Necker hospital, Paris) and Philippe Henchoz (Service de Pathologie Clinique, Hôpitaux Universitaires de Genève, Genève) for their technical assistance.

## FUNDING

The research leading to these results has received funding from the VERUM foundation and “Investissements d’avenir” ANR-11-INBS-0011 – NeurATRIS: Translational Research Infrastructure for Biotherapies in Neurosciences. This work was funded by the Programme Hospitalier de Recherche Clinique (PHRC) FTLD-exome (to I.L.B., promotion by Assistance Publique – Hôpitaux de Paris) and PHRC Predict-PGRN (to I.L.B., promotion by Assistance Publique – Hôpitaux de Paris).

## DECLARATION OF INTERESTS

The authors disclose no conflicts of interest.

## CONTRIBUTIONS

| Name | Location | Role | Contribution |
|---|---|---|---|
| Vincent Huin, MD, PhD | Sorbonne Université | Author | Major role in the acquisition of data, interpreted the data, drafted the manuscript for intellectual content |
| Mathieu Barbier, PhD | Sorbonne Université | Author | Role in the acquisition of data, drafted the manuscript for intellectual content |
| Armand Bottani, MD | Hôpitaux universitaires de Genève | Author | Acquisition of data |
| Johannes Alexander Lobrinus, MD | Hôpitaux universitaires de Genève | Author | Acquisition of data |
| Fabienne Clot, PhD | APHP | Author | Acquisition of data |
| Foudil Lamari, PharmD, PhD | APHP | Author | Acquisition of data |
| Laureen Chat | APHP | Author | Acquisition of data |
| Benoît Rucheton, PharmD | APHP | Author | Acquisition of data |
| Frédérique Fluchere, MD | Université d'Aix-Marseille | Author | Acquisition of data |
| Stéphane Auvin, MD, | APHP | Author | Acquisition of data |
| PhD | | | |
| Peter Myers, MD | Cabinet Médical, Genève | Author | Acquisition of data |
| Antoinette Bernabe Gelot, MD, PhD | APHP, Université d'Aix-Marseille | Author | Acquisition of data |
| Agnès Canuzat, PhD | Sorbonne Université | Author | Acquisition of data |
| Catherine Caillaud, MD, PhD | APHP | Author | Acquisition of data |
| Ludmila Jornéa, BSc | Sorbonne Université | Author | Acquisition of data |
| Sylvie Forlani, PhD | Sorbonne Université | Author | Acquisition of data |
| Dario Saracino, MD | Sorbonne Université | Author | Acquisition of data |
| Charles Duyckaerts, MD, PhD | APHP | Author | Acquisition of data |
| Alexis Brice, MD | Sorbonne Université | Author | Revised the manuscript for intellectual content |
| Alexandra Durr, MD, PhD | Sorbonne Université | Author | Major role in the acquisition of data, revised the manuscript for intellectual content |
| Isabelle Le Ber, MD, PhD | Sorbonne Université | Author | Major role in the acquisition of data, revised the manuscript for intellectual content |

## REFERENCES

Ahmed Z, Sheng H, Xu Y-F, Lin W-L, Innes AE, Gass J, et al. Accelerated lipofuscinosis and ubiquitination in granulin knockout mice suggest a role for progranulin in successful aging. Am J Pathol 2010; 177: 311–24.

Almeida MR, Baldeiras I, Ribeiro MH, Santiago B, Machado C, Massano J, et al. Progranulin peripheral levels as a screening tool for the identification of subjects with progranulin mutations in a Portuguese cohort. Neurodegener Dis 2014; 13: 214–23.

Almeida MR, Macário MC, Ramos L, Baldeiras I, Ribeiro MH, Santana I. Portuguese family with the co-occurrence of frontotemporal lobar degeneration and neuronal ceroid lipofuscinosis phenotypes due to progranulin gene mutation. Neurobiol Aging 2016; 41: 200.e1–e5.

Anderson GW, Goebel HH, Simonati A. Human pathology in NCL. Biochim Biophys Acta (BBA) 2013; 1832: 1807–26.

Arsov T, Smith KR, Damiano J, Franceschetti S, Canafoglia L, Bromhead CJ, et al. Kufs disease, the major adult form of neuronal ceroid lipofuscinosis, aused by mutations in CLN6. Am J Hum Genet 2011; 88: 566–573.

Baker M, Mackenzie IR, Pickering-Brown SM, Gass J, Rademakers R, Lindholm C, et al. Mutations in progranulin cause tau-negative frontotemporal dementia linked to chromosome 17. Nature 2006; 442: 916–9.

Benussi L, Ghidoni R, Pegoiani E, Moretti DV, Zanetti O, Binetti G. Progranulin Leu271LeufsX10 is one of the most common FTLD and CBS associated mutations worldwide. Neurobiol Dis 2009; 33: 379–85.

Brandmeir NJ, Geser F, Kwong LK, Zimmerman E, Qian J, Lee VMY, et al. Severe subcortical TDP-43 pathology in sporadic frontotemporal lobar degeneration with motor neuron disease. Acta Neuropathol 2008; 115: 123–31.

Calvi A, Cioffi SMG, Caffarra P, Fenoglio C, Serpente M, Pietroboni AM, et al. The novel GRN g.1159_1160delTG mutation is associated with behavioral variant frontotemporal dementia. J Alzheimer’s Dis 2015; 44: 277–82.

Canafoglia L, Morbin M, Scaioli V, Pareyson D, D’Incerti L, Fugnanesi V, et al. Recurrent generalized seizures, visual loss, and palinopsia as phenotypic features of neuronal ceroid lipofuscinosis due to progranulin gene mutation. Epilepsia 2014; 55: e56–9.

Coutelier M, Burglen L, Mundwiller E, Abada-Bendib M, Rodriguez D, Chantot-Bastaraud S, et al. GRID2 mutations span from congenital to mild adult-onset cerebellar ataxia. Neurology 2015; 84: 1751–9.

Cruts M, Gijselinck I, van der Zee J, Engelborghs S, Wils H, Pirici D, et al. Null mutations in progranulin cause ubiquitin-positive frontotemporal dementia linked to chromosome 17q21. Nature 2006; 442: 920–4.

Faber I, Prota JRM, Martinez ARM, Lopes-Cendes I, França MC. A new phenotype associated with homozygous GRN mutations: complicated spastic paraplegia. Eur J Neurol 2017; 24: e3–4.

Finch N, Baker M, Crook R, Swanson K, Kuntz K, Surtees R, et al. Plasma progranulin levels predict progranulin mutation status in frontotemporal dementia patients and asymptomatic family members. Brain 2009; 132: 583–91.

Finch N, Carrasquillo MM, Baker M, Rutherford NJ, Coppola G, Dejesus-Hernandez M, et al. TMEM106B regulates progranulin levels and the penetrance of FTLD in GRN mutation carriers. Neurology 2011; 76: 467–74.

Galimberti D, Bertram K, Formica A, Fenoglio C, Cioffi SMG, Arighi A, et al. Plasma screening for progranulin mutations in patients with progressive supranuclear palsy and corticobasal syndromes. J Alzheimer's Dis 2016; 53: 445–9.

Galimberti D, Fumagalli GG, Fenoglio C, Cioffi SMG, Arighi A, Serpente M, et al. Progranulin plasma levels predict the presence of GRN mutations in asymptomatic subjects and do not correlate with brain atrophy: results from the GENFI study. Neurobiol Aging 2018; 62: 245.e9–e12.

Ghidoni R, Benussi L, Glionna M, Franzoni M, Binetti G. Low plasma progranulin levels predict progranulin mutations in frontotemporal lobar degeneration. Neurology 2008; 71: 1235–9.

Ghidoni R, Stoppani E, Rossi G, Piccoli E, Albertini V, Paterlini A, et al. Optimal plasma progranulin cutoff value for predicting null progranulin mutations in neurodegenerative diseases: a multicenter Italian study. Neurodegener Dis 2012; 9: 121–7.

Gorno-Tempini ML, Hillis AE, Weintraub S, Kertesz A, Mendez M, Cappa SF, et al. Classification of primary progressive aphasia and its variants. Neurology 2011; 76: 1006–14.

Groh M, Albulescu LO, Cristini A, Gromak N. Senataxin: genome guardian at the interface of transcription and neurodegeneration. J Mol Biol 2017; 429: 3181–95.

Hu F, Padukkavidana T, Vægter CB, Brady OA, Zheng Y, Mackenzie IR, et al. Sortilin-mediated endocytosis determines levels of the frontotemporal dementia protein, progranulin. Neuron 2010; 68: 654–67.

Jian J, Tian Q-Y, Hettinghouse A, Zhao S, Liu H, Wei J, et al. Progranulin recruits HSP70 to b-glucocerebrosidase and is therapeutic against Gaucher disease. EBioMedicine 2016; 13: 212–24.

Kamate M, Detroja M, Hattiholi V. Neuronal ceroid lipofuscinosis type-11 in an adolescent. Brain Dev 2019; 41: 542–5.

Lattante S, Le Ber I, Galimberti D, Serpente M, Rivaud-Péchoux S, Camuzat A, et al. Defining the association of TMEM106B variants among frontotemporal lobar degeneration patients with GRN mutations and C9orf72 repeat expansions. Neurobiol Aging 2014; 35: 2658.e1–e5.

Lauronen L, Munroe PB, Jarvela I, Autti T, Mitchison HM, O'Rawe AM, et al. Delayed classic and protracted phenotypes of compound heterozygous juvenile neuronal ceroid lipofuscinosis. Neurology 1999; 52: 360–360.

Le Ber I, Camuzat A, Guillot-Noel L, Hannequin D, Lacomblez L; The French Research Network on FTLD/FTLD-ALS et al. C9ORF72 repeat expansions in the frontotemporal dementias spectrum of diseases: a flow-chart for genetic testing. J Alzheimer's Dis 2013; 34: 485–99.

Le Ber I, Camuzat A, Hannequin D, Pasquier F, Guedj E, Rovelet-Lecrux A, et al. Phenotype variability in progranulin mutation carriers: a clinical, neuropsychological, imaging and genetic study. Brain 2008; 131: 732–46.

Lee EB, Lee VM-Y, Trojanowski JQ. Gains or losses: molecular mechanisms of TDP43-mediated neurodegeneration. Nat Rev Neurosci 2012; 13: 38–50.

Mackenzie IRA, Neumann M, Baborie A, Sampathu DM, Du Plessis D, Jaros E, et al. A harmonized classification system for FTLD-TDP pathology. Acta Neuropathol. 2011; 122: 111–3.

Martens LH, Zhang J, Barmada SJ, Zhou P, Kamiya S, Sun B, et al. Progranulin deficiency promotes neuroinflammation and neuron loss following toxin-induced injury. J Clin Invest 2012; 122: 3955–9.

Mole SE, Anderson G, Band HA, Berkovic SF, Cooper JD, Kleine Holthaus S-M, et al. Clinical challenges and future therapeutic approaches for neuronal ceroid lipofuscinosis. Lancet Neurol 2019; 18: 107–16.

Mole SE, Williams RE, Goebel HH, editors. The neuronal ceroid lipofuscinoses (Batten disease). 2 edn. Oxford: Oxford University Press; 2011. p. 35–49.

Mukherjee AB, Appu AP, Sadhukhan T, Casey S, Mondal A, Zhang Z, et al. Emerging new roles of the lysosome and neuronal ceroid lipofuscinoses. Mol Neurodegener 2019; 14: 4.

Perry DC, Lehmann M, Yokoyama JS, Karydas A, Lee JJ, Coppola G, et al. Progranulin mutations as risk factors for Alzheimer disease. JAMA Neurol 2013; 70: 774–8.

Pires C, Coelho M, Valadas A, Barroso C, Pimentel J, Martins M, et al. Phenotypic variability of familial and sporadic progranulin p.Gln257Profs_27 mutation. J Alzheimer's Dis 2013; 37: 335–42.

Pottier C, Zhou X, Perkerson RB, Baker M, Jenkins GD, Serie DJ, et al. Potential genetic modifiers of disease risk and age at onset in patients with frontotemporal lobar degeneration and GRN mutations: a genome-wide association study. Lancet Neurol 2018; 17: 548–58.

Rademakers R, Baker M, Gass J, Adamson J, Huey ED, Momeni P, et al. Phenotypic variability associated with progranulin haploinsufficiency in patients with the common 1477C–4T (Arg493X) mutation: an international initiative. Lancet Neurol 2007; 6: 857–68.

Rascovsky K, Hodges JR, Knopman D, Mendez MF, Kramer JH, Neuhaus J, et al. Sensitivity of revised diagnostic criteria for the behavioural variant of frontotemporal dementia. Brain 2011; 134: 2456–77.

Seilhean D, Le Ber I, Sarazin M, Lacomblez L, Millecamps S, Salachas F, et al. Fronto-temporal lobar degeneration: neuropathology in 60 cases. J Neural Transm 2011; 118: 753–64.

Sibley CR, Blazquez L, Ule J. Lessons from non-canonical splicing. Nat Rev Genet 2016; 17: 407–21.

Smith KR, Dahl H-HM, Canafoglia L, Andermann E, Damiano J, Morbin M, et al. Cathepsin F mutations cause Type B Kufs disease, an adult-onset neuronal ceroid lipofuscinosis. Hum Mol Genet 2013; 22: 1417–23.

Smith KR, Damiano J, Franceschetti S, Carpenter S, Canafoglia L, Morbin M, et al. Strikingly different clinicopathological phenotypes determined by progranulin-mutation dosage. Am J Hum Genet 2012; 90: 1102–7.

Van Damme P, Van Hoecke A, Lambrechts D, Vanacker P, Bogaert E, van Swieten J, et al. Progranulin functions as a neurotrophic factor to regulate neurite outgrowth and enhance neuronal survival. J Cell Biol 2008; 181: 37–41.

Van Deerlin VM, Sleiman PMA, Martinez-Lage M, Chen-Plotkin A, Wang L-S, Graff-Radford NR, et al. Common variants at 7p21 are associated with frontotemporal lobar degeneration with TDP-43 inclusions. Nat Genet 2010; 42: 234–9.

van der Zee J, Mariën P, Crols R, Van Mossevelde S, Dillen L, Perrone F, et al. Mutated CTSF in adult-onset neuronal ceroid lipofuscinosis and FTD. Neurol Genet 2016; 2: e102.

van Diggelen OP, Thobois S, Tilikete C, Zabot MT, Keulemans JL, van Bunderen PA, et al. Adult neuronal ceroid lipofuscinosis with palmitoyl-protein thioesterase deficiency: first adult-onset patients of a childhood disease. Ann Neurol 2001; 50: 269–72.

Vatsavayai SC, Yoon SJ, Gardner RC, Gendron TF, Vargas JNS, Trujillo A, et al. Timing and significance of pathological features in C9orf72 expansion-associated frontotemporal dementia. Brain 2016; 139: 3202–16.

Ward ME, Chen R, Huang H-Y, Ludwig C, Telpoukhovskaia M, Taubes A, et al. Individuals with progranulin haploinsufficiency exhibit features of neuronal ceroid lipofuscinosis. Sci Transl Med 2017; 9: eaah5642.

Xin W, Mullen TE, Kiely R, Min J, Feng X, Cao Y, et al. CLN5 mutations are frequent in juvenile and late-onset non-Finnish patients with NCL. Neurology 2010; 74: 565–71.

Zhou X, Paushter DH, Feng T, Pardon CM, Mendoza CS, Hu F. Regulation of cathepsin D activity by the FTLD protein progranulin. Acta Neuropathol 2017b; 134: 151–3.

Zhou X, Paushter DH, Feng T, Sun L, Reinheckel T, Hu F. Lysosomal processing of progranulin. Mol Neurodegener 2017a; 12: 62.

| **Study** | Smith *et al.*, 2012<br>Canafoglia *et al.*, 2014 | | Almeida *et al.*, 2016 | Faber *et al.*, 2017 | Kamate *et al.*, 2019 | This study | | | | | |
|---|---|---|---|---|---|---|---|---|---|---|---|
| **Family** | 1 | | 2 | 3 | 4 | FTD-1042 | FTDP-N12/1611 | | AAR-427 | | NCL-001 |
| **Subject** | Proband | Sister | Proband | Proband | Proband | Proband | Proband | Brother | Proband | Sister | Proband |
| **Sex** | Male | Female | Female | Female | Female | Female | Male | Male | Female | Female | Female |
| **Origin** | Italy | | Portugal | Brazil | India | France | France | | Portugal | | Pakistan |
| **Pathogenic variant**[1] | c.813_816del<br>p.(Thr272Serfs*10) | | c.900_901dup<br>p.(Ser301Cysfs*61) | c.768_769dup<br>p.(Gln257Profs*27) | c.912G>A<br>p.(Trp304*) | c.709-3C>G<br>p.? | c.443_444del<br>p.(Gly148Valfs*11) | | c.768_769dup<br>p.(Gln257Profs*27) | | c.1A>T<br>p.(Met1?) |
| **Familal history (parents)** | Late dementia | | FTD | None | None | None | Dementia, Parkinson | | None | | None |
| **Age of onset (years)** | 22 | 23 | 25 | 21 | 13 | 56 | 44 | 18 | 16 | 12 | 7 |
| **Disease duration (years)** | 7[4] | 3[4] | 9[4] | 4[4] | 1[4] | 2[4] | 22[3] | 46[3] | 11[3] | 24[4] | 7[4] |
| **Dominant phenotype** | CLN11 | CLN11 | CLN11 | Spastic ataxia | CLN11 | bvFTD | RP-FTDP | RP-FTDP | CLN11 | CLN11 | CLN11 |
| **First sign** | Visual loss | Seizure | Visual loss | Gait impairment | Seizure | Visual hallucinations | Visual loss | Visual loss | Seizure | Seizure | Seizure |
| **Symptoms** | | | | | | | | | | | |
| Visual loss | ++ | + | +++ | n.a. | - | - | +++ | +++ | +++ | +++ | ++ |
| Seizure | ++ | ++ | - | + | +++ | - | - | - | +++ | +++ | +++ |
| Cerebellar ataxia | + | + | + | +++ | + | - | - | - | +++ | +++ | + |
| Pyramidal syndrome | - | - | - | + | - | + | - | - | - | - | - |
| Cognitive deterioration | + | - | - | ++ | + | +++ | +++ | +++ | ++ | + | ++ |
| Behavourial changes | - | - | - | ++ | - | +++ | +++ | +++ | ++ | - | - |
| Akinetic-rigid syndrome | - | - | - | - | - | + | +++ | +++ | - | - | - |
| Visual hallucinations | - | - | - | - | - | + | + | - | + | + | - |
| **PGRN dosage (µg/L)** | Undectectable | Undectectable | Undectectable | n.a. | n.a. | 30-39 | Undectectable | Undectectable. | n.a. | Undectectable | Undectectable |
| **Brain imaging** | | | | | | | | | | | |
| Cerebellar involvement[2] | ++ | +++ | +++ | +++ | +++ | - | - | - | +++ | ++ | ++ |
| FT involvement[2] | - | - | - | - | - | ++ | +++ | + | + | + | + |
| **Pathology examination** | Skin, lymphocytes | Lymphocytes | Skin | n.a. | n.a. | n.a. | n.a. | n.a. | Brain, lymphocytes | Lymphocytes | Skin |
| Cytoplasmic vacuoles | + | - | n.a. | n.a. | n.a. | n.a. | n.a. | n.a. | + | + | + |
| Curvilinear profiles | + | + | n.a. | n.a. | n.a. | n.a. | n.a. | n.a. | + | n.a. | + |
| Fingerprint profiles | + | - | n.a. | n.a. | n.a. | n.a. | n.a. | n.a. | + | n.a. | - |
| Lipofuscin inclusions | n.a. | n.a. | + | n.a. | n.a. | n.a. | n.a. | n.a. | + | n.a. | n.a. |

**Table 1. Clinical features of four published cases and six novel cases reported in this study carrying biallelic *GRN* mutations.**

-, absent; +, mild; ++, moderate, +++, severe; n.a., not available. FTD: frontotemporal dementia; FTDP: frontotemporal dementia and parkinsonism; RP: retinitis pigmentosa. [1] Mutations at the homozygous state in all the cases, [2] Atrophy or hypometabolism; [3] Disease duration at death; [4] Disease duration at last examination.

## FIGURE LEGENDS

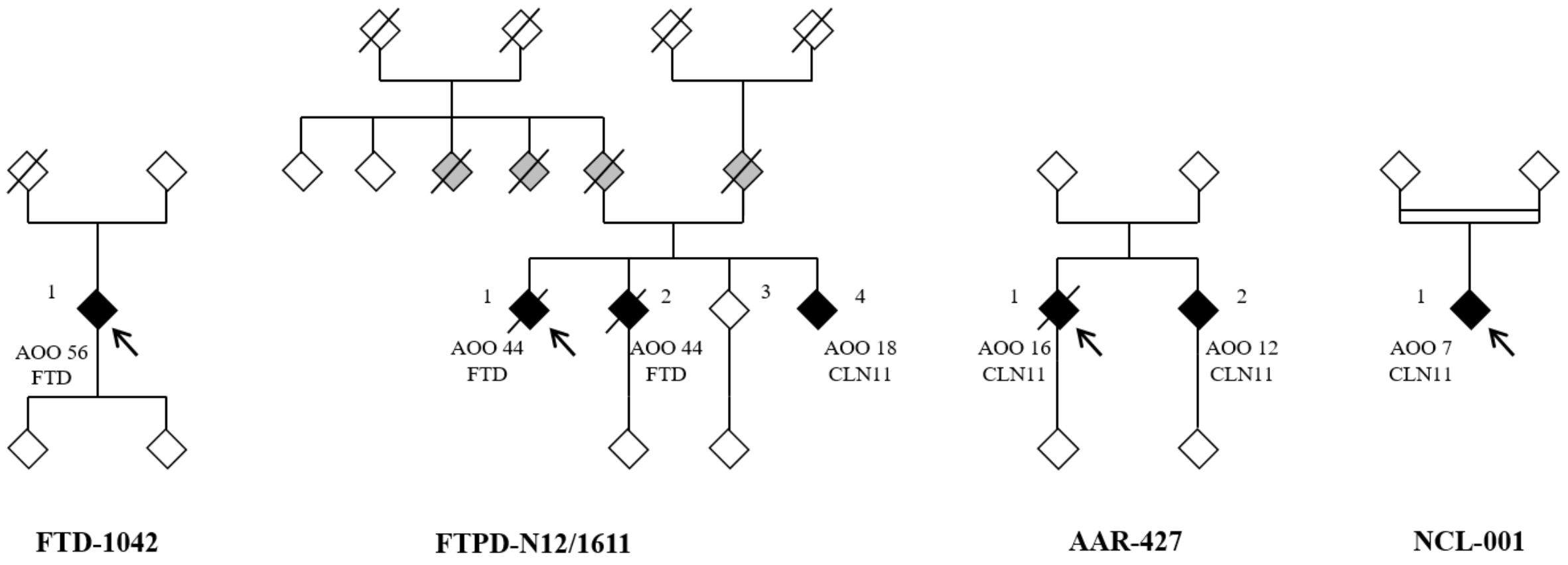

**Figure 1. Family trees of four families described in this study.**

The probands are indicated by an arrow. Filled black symbols denote members clinically affected by FTD, RP-FTD (retinitis pigmentosa and FTD) or CLN11 phenotypes. Grey filled symbols denote members clinically affected by other neurodegenerative disorders or unspecified dementia. Numbers indicate the age of onset (AOO) in years. Open symbols indicate unaffected individuals. Forward slash indicates deceased individuals. Family trees have been anonymized for confidentiality.

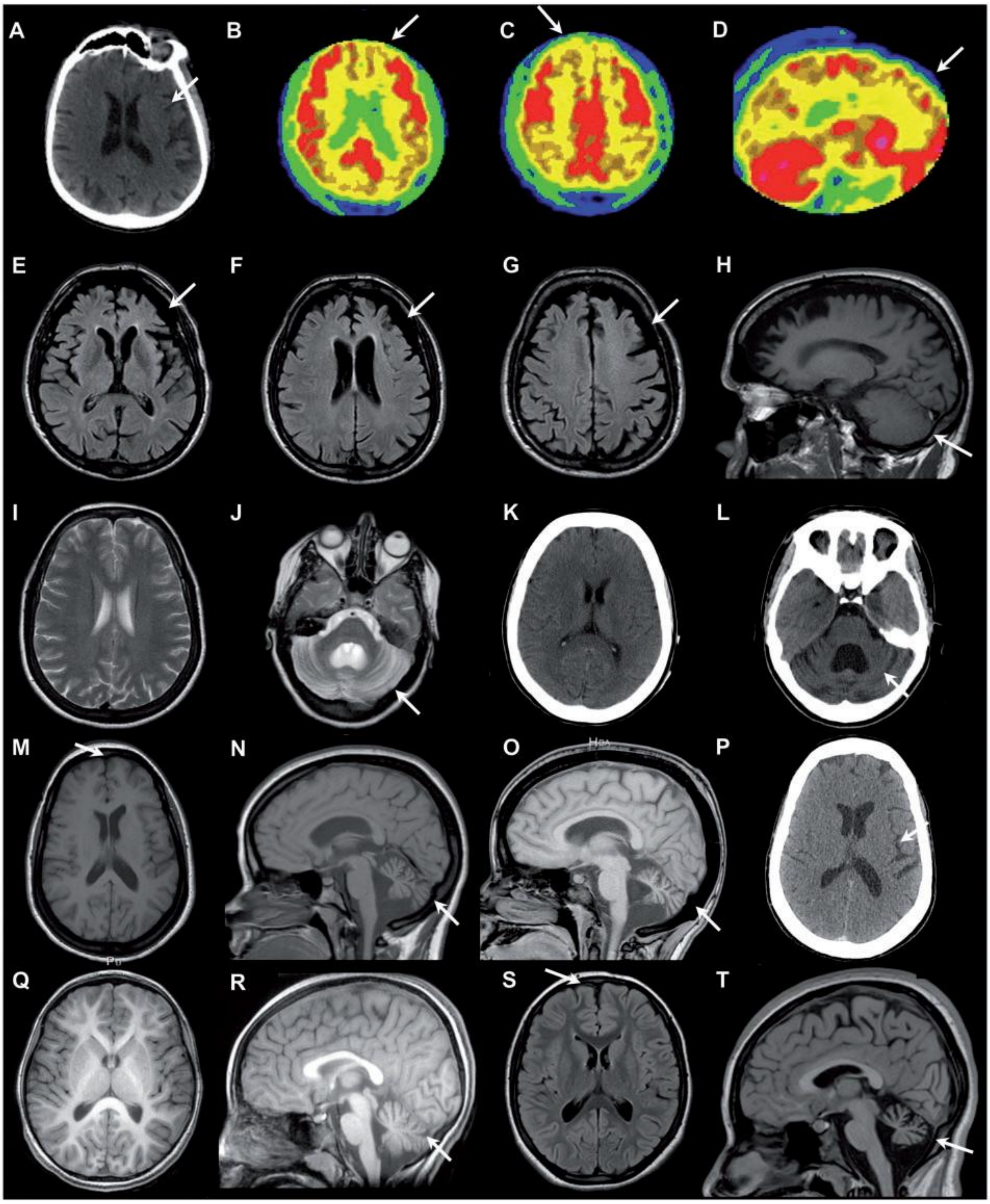

**Figure 2. Neuroimaging characteristics in homozygous *GRN* carriers.**

**A–D.** Patient 1 from Family FTDP-N12/1611 (proband). **A.** Brain CT scan at age 63, showing left perisylvian atrophy (arrow). **B–D.** FDG-PET scan at age 63, axial (B and C) and sagittal (D) sections, showing hypometabolism in the cingular and mesial prefrontal cortex (arrows), extending to the parieto-occipital regions. Cerebellum metabolism was normal. **E–H.** Brain MRI scan of Patient 2 of family FTDP-N12/1611. **E–G.** FLAIR axial sections showing frontotemporal atrophy (arrow). **H.** T1 sagittal sequence demonstrating the absence of cerebellar atrophy (arrow). **I–L.** Brain MRI and CT scan of Patient 1 (proband) from Family AAR-427. (I and J) Brain MRI, T2 axial sections at age 22, showing discrete enlargement of brain sulci in the frontal lobes and severe cerebellar atrophy (arrow). (K and L) Brain CT scan at age 27, showing progression of cerebellar atrophy (arrow) and discrete enlargement of brain sulci in the frontal lobes. **M–P.** Brain MRI and CT scan of Patient 2 from Family AAR-427. (M–O) Brain MRI, T1 axial (M) and sagittal (N) sections, showing severe cerebellar atrophy (arrows) at age 22, with progression at age 29 (O). (P) Brain CT scan at age 34, showing moderate bilateral frontal and peri-sylvian atrophy (arrow). **Q–T.** Brain MRI of the proband of family NCL-001. (Q) T1 axial section at age 8, showing no significant cortical atrophy. (R) T1 sagittal section at age 8, showing cerebellar atrophy (arrow). (S) FLAIR axial section at age 12, showing moderate frontal atrophy (arrow). (T) T1 sagittal section at age 12, showing progression of cerebellar atrophy (arrow).

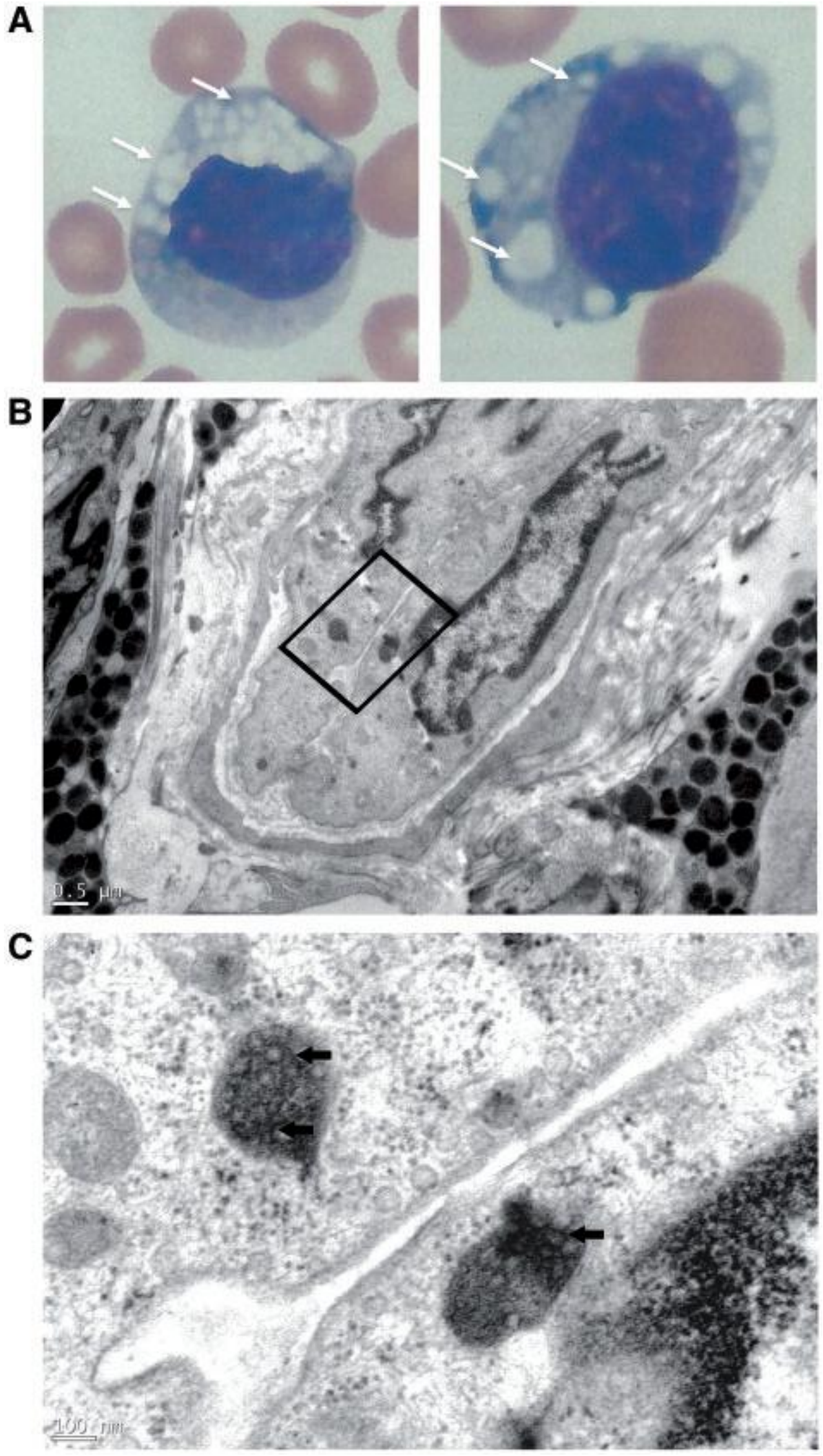

**Figure 3. Histological and ultrastructural images of neuronal ceroid lipofuscinosis-11 in skin and lymphocytes, in families NCL-001 and AAR-427.**

**A** Photomicrograph of vacuolated lymphocytes. May-Grunwald-Giemsa stained blood film of the proband (patient 1) from family AAR-427 (left) and patient 2 (right) showing lymphocytes with many large bold vacuoles. **B.** Electron microscopy of skin biopsy of the proband of family NCL-001 showing rare inclusions in the cytoplasm of endothelial cells (initial magnification x 2500). These vesicles, of which the content is electron dense, are surrounded by a simple membrane. **C.** At a magnification of x 15 000, ultrastructural analysis reveals these vesicles to be filled with membrane formations wound on themselves, with a morphology similar to that of curvilinear profiles, except that they are more compact and form a nearly complete circle.

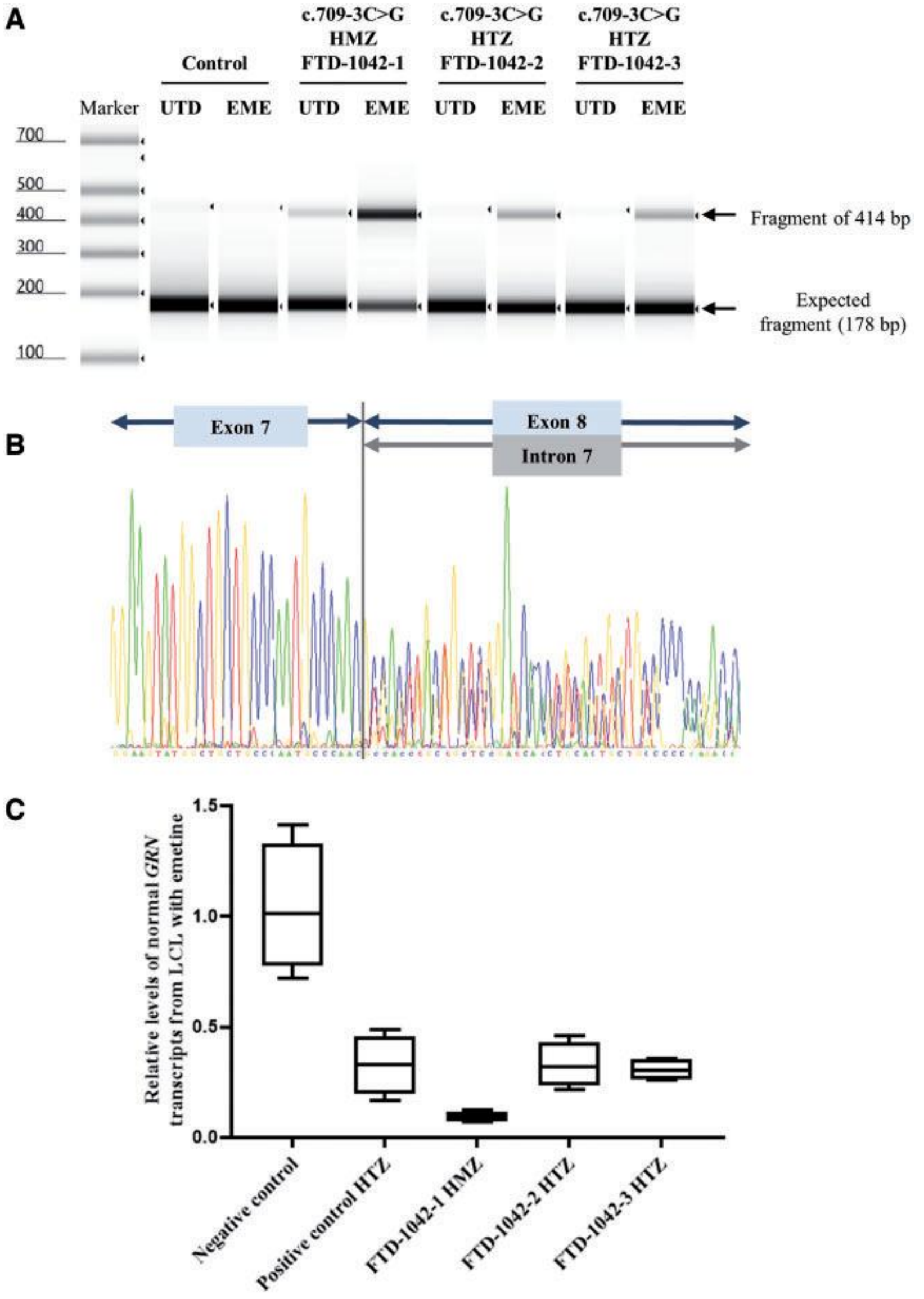

**Figure 4. Effect of the mutation c.709-3C>G on RNA (family FTD-1042).**

**A.** RT-PCR analysis of GRN gene expression in lymphoblast cell lines from a healthy control, FTD-1042-1 (homozygous carrier), and FTD-1042-2 and 3 (heterozygous carriers). Lymphoblast cell lines were previously untreated (UTD) or treated with emetine (EME). Two transcripts were detected: the expected fragment (178 bp) and a larger fragment (414 bp) (analysis not quantitative). **B.** Sequencing chromatogram of RT-PCR from lymphoblast cell lines of FTD-1042-1 treated with EME. The sequence shows the retention of full-length intron 7 of 236 bp. (C) Box plots representing the relative values of normal GRN transcripts from lymphoblast cell lines (LCL) treated with EME from the homozygous patient, heterozygous children, positive control carrying a heterozygous c.813_816del, p.(Thr272Serfs*10) mutation, and non-mutated controls. HMZ = homozygous; HTZ = heterozygous.

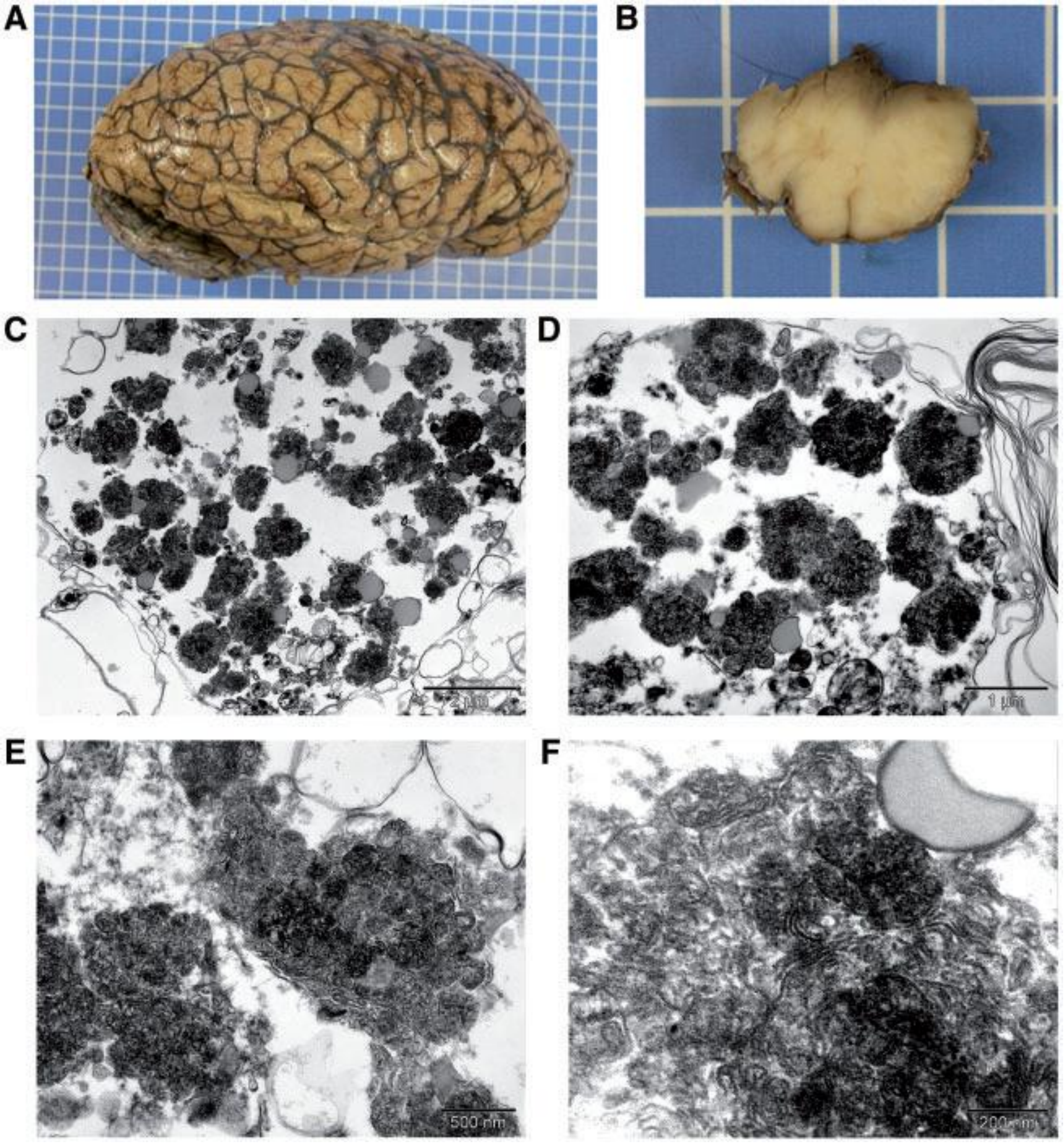

**Figure 5. Neuropathology of patient 1 from family AAR-427: Macroscopy and electronic microscopy.**

**A.** Brain macroscopy showing cerebellar atrophy. No marked frontal atrophy was observed. **B.** Brain macroscopy of medulla oblongata showing pallor of olivary nuclei, highlighting neuronal loss. **C–F.** Electron microscopy of temporal cortex (C–E) and occipital cortex (F) neurons, showing numerous lysosomal deposits (C and D), with curvilinear-like profiles (E and F) and few small fingerprint profiles (F). Scale bars in C = 2 µm; D = 1 µm; E = 500 nm; F = 200 nm

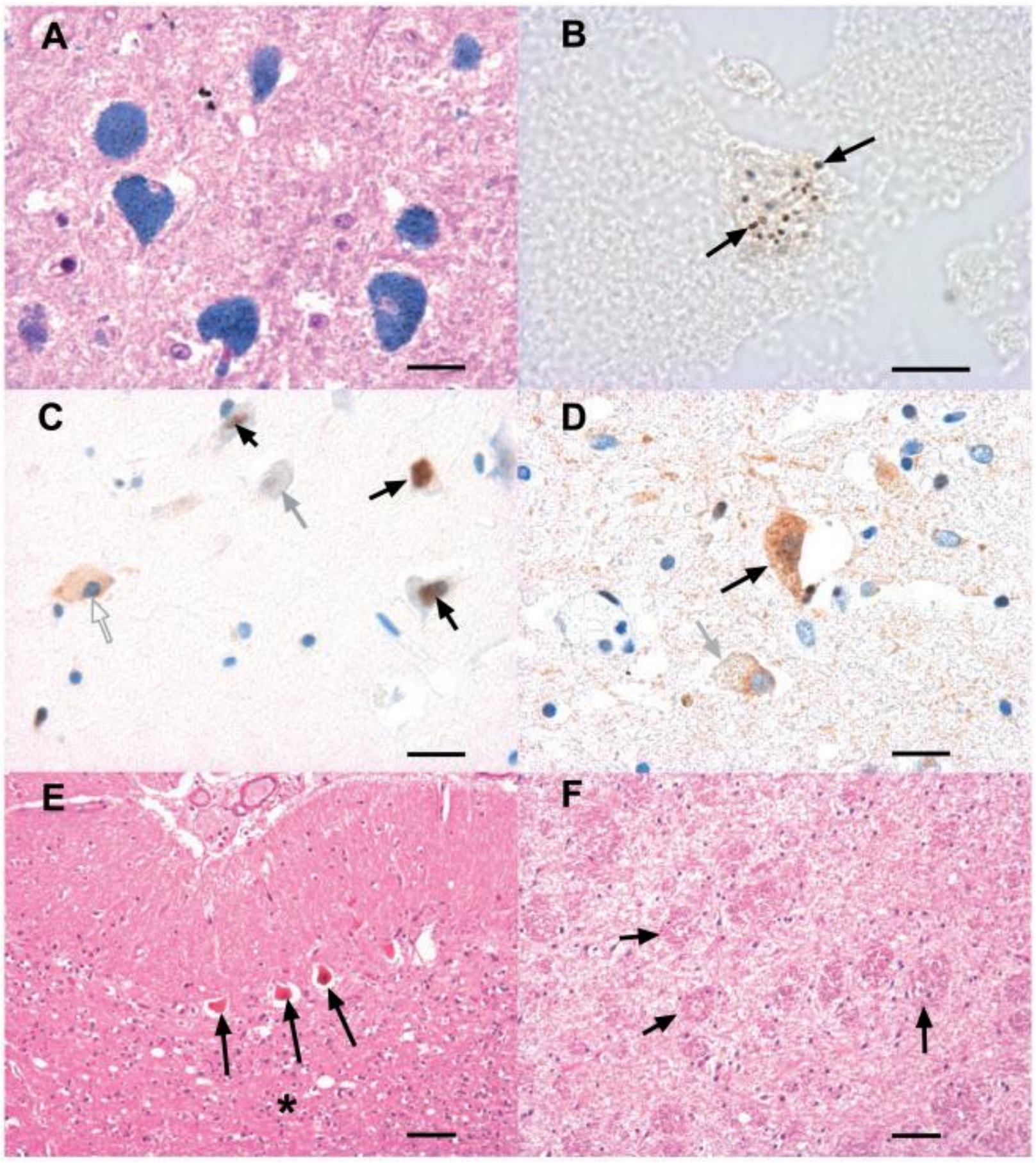

**Figure 6 Neuropathology of Patient 1 from Family AAR-427: immunochemistry.**

**A.** Ballooned neurons in the CA4 sector of the hippocampus. Lipofuscin granules, filling the cell body of the neurons, are stained in blue by Luxol fast blue. **B**. Formalin pigment in unstained neurons. The pigments are localized in lipofuscin-rich neurons. **C.** TDP-43 immunohistochemistry in the subiculum. Filled arrows: normal nuclear localization of TDP-43 immunoreactivity. Open arrow: TDP-43 immunoreactivity is abnormally localized to the cell body of a neuron; the nucleus is negative. Grey arrow: both the cell body and the nucleus are immunonegative. **D**. TDP-43 immunohistochemistry in the thalamus. Black arrow: dense immunoreactivity in the cell body of a neuron. Grey arrow: in that neuron, the nucleus is immuno-negative. A ring of cytoplasmic immunoreactivity is seen around the nucleus. **E.** Cerebellar cortex. Haematoxylin and eosin stain. Three Purkinje cells are visible, which are filled with lipofuscin. There is severe neuronal loss in the granular layer (asterisk), where glomeruli are no longer visible. **F.** Dentate nucleus of the cerebellum. Haematoxylin and eosin stain. Grumose degeneration. No neuron is visible. Numerous aggregates of axons (black arrows), some indicated by a black arrow. Scale bars in A, C and D = 20 µm; B = 10 µm; E and F = 50 µm.

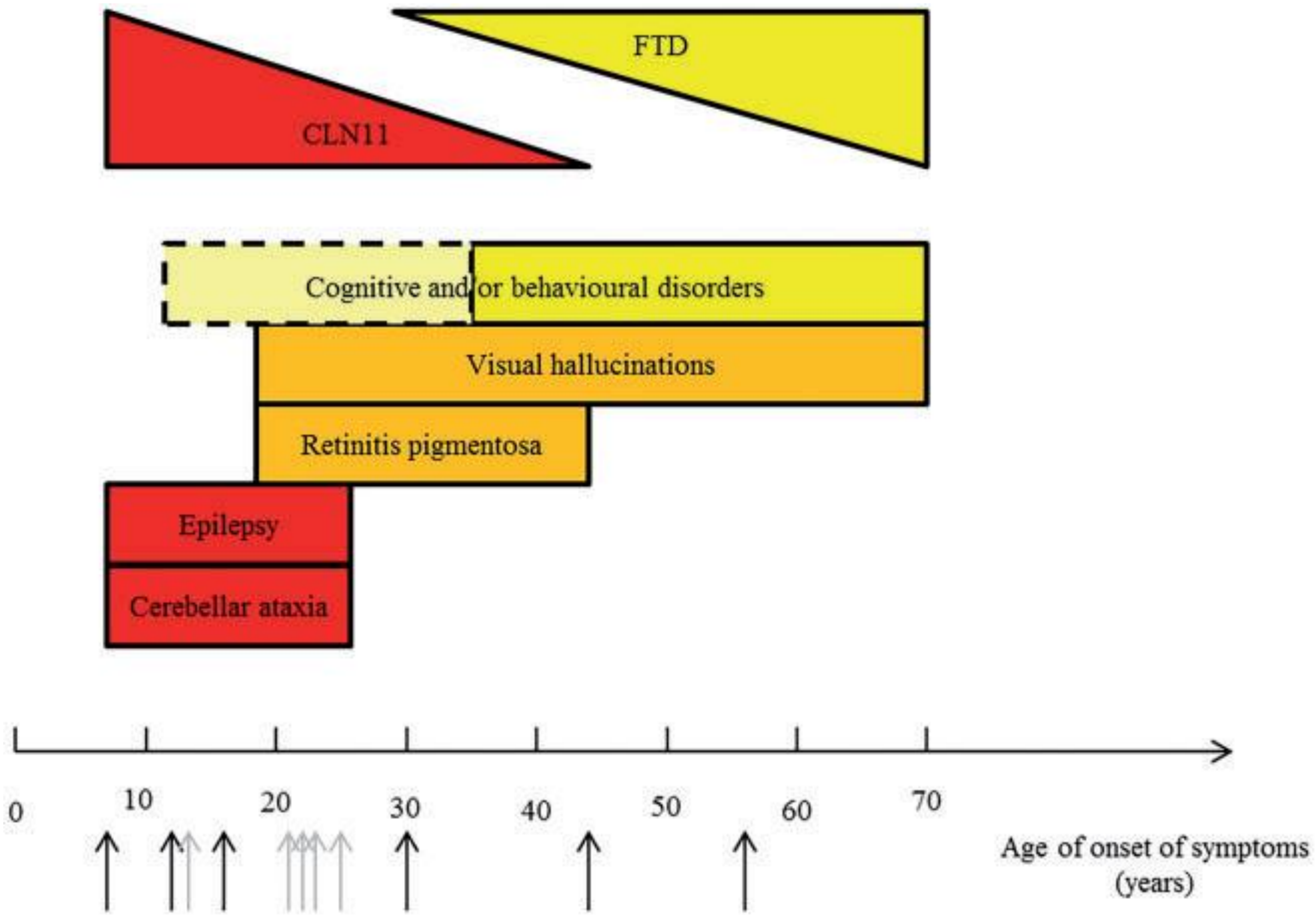

**Figure 7. Clinical continuum of homozygous *GRN*-related disorders.**

The X axis represents the age at onset of symptoms. Arrows indicate the age at onset of the patients with homozygous *GRN* mutation reported here (in black) or in previous reports (in grey).

**SUPLEMENTARY DATA**

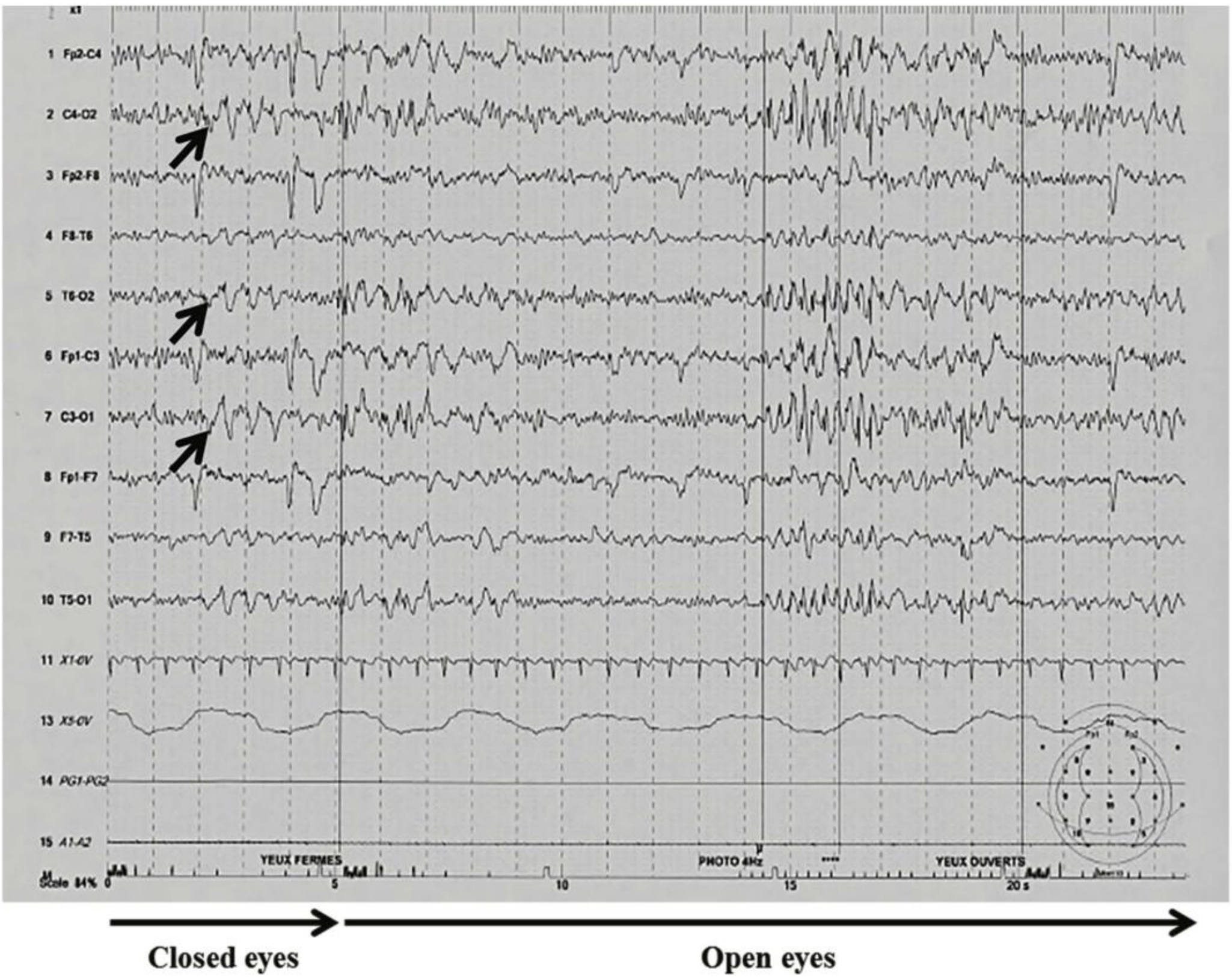

**Supplemental Figure S1. EEG traces of the proband from family NCL-001.**

EEG traces, showing spikes as photoparoxysmal response in low frequency, predominating in the posterior regions (arrows).

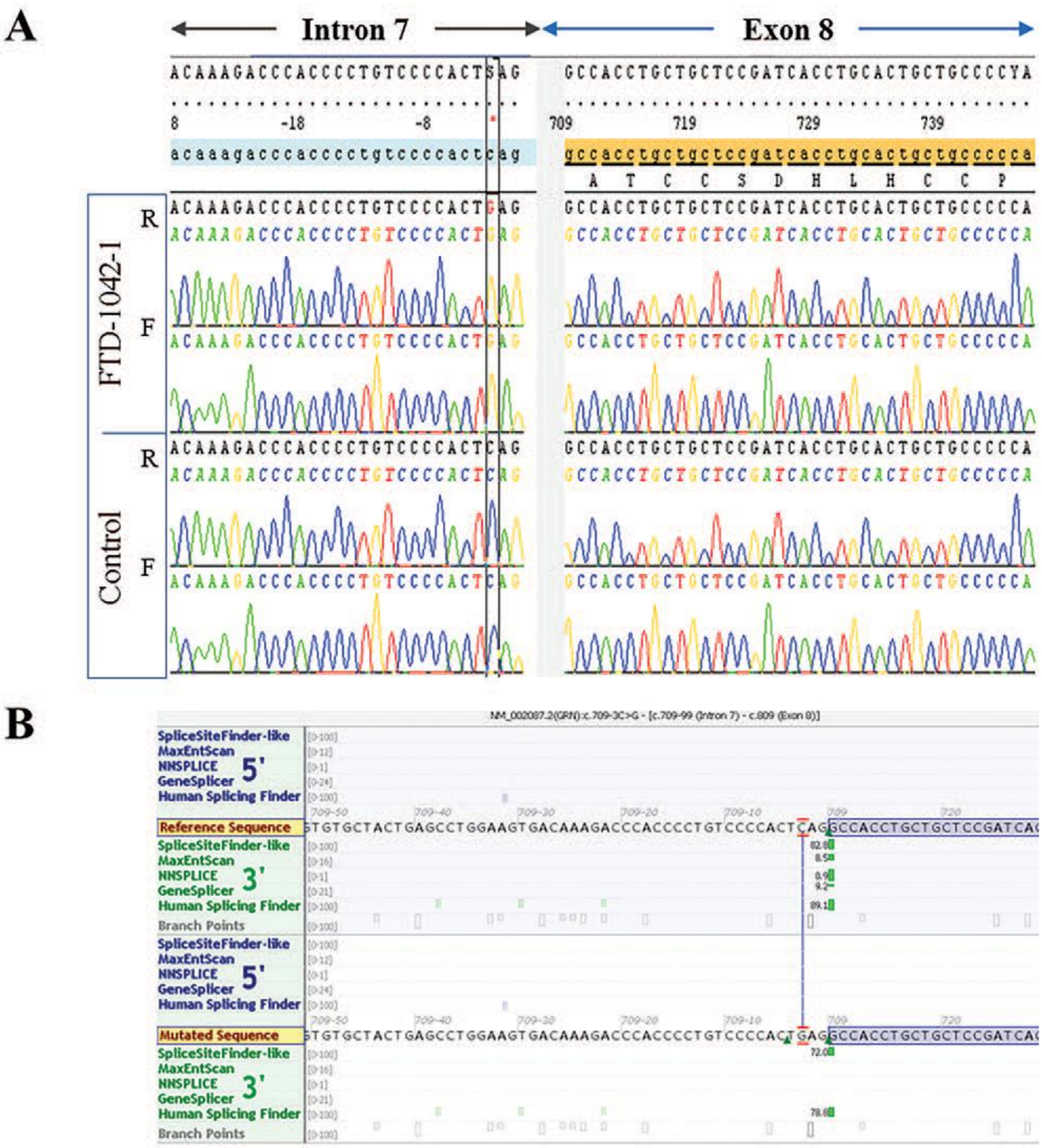

**Supplemental Figure S2. Genomic DNA sequence of the intron7-exon 8 junction and predicted impact on the acceptor splice site.**

**A.** The genomic DNA sequence showed the c.709-3C>G mutation at the homozygous state in the proband of family FTD-1042, and in healthy control. R: reverse strand, F: forward strand. **B.** Predicted impact of the mutation on the acceptor splice site using *in silico* GeneSplicer, NNSPLICE, MaxEntScan and SpliceSiteFinder-like and HSFsoftwares (from Alamut).

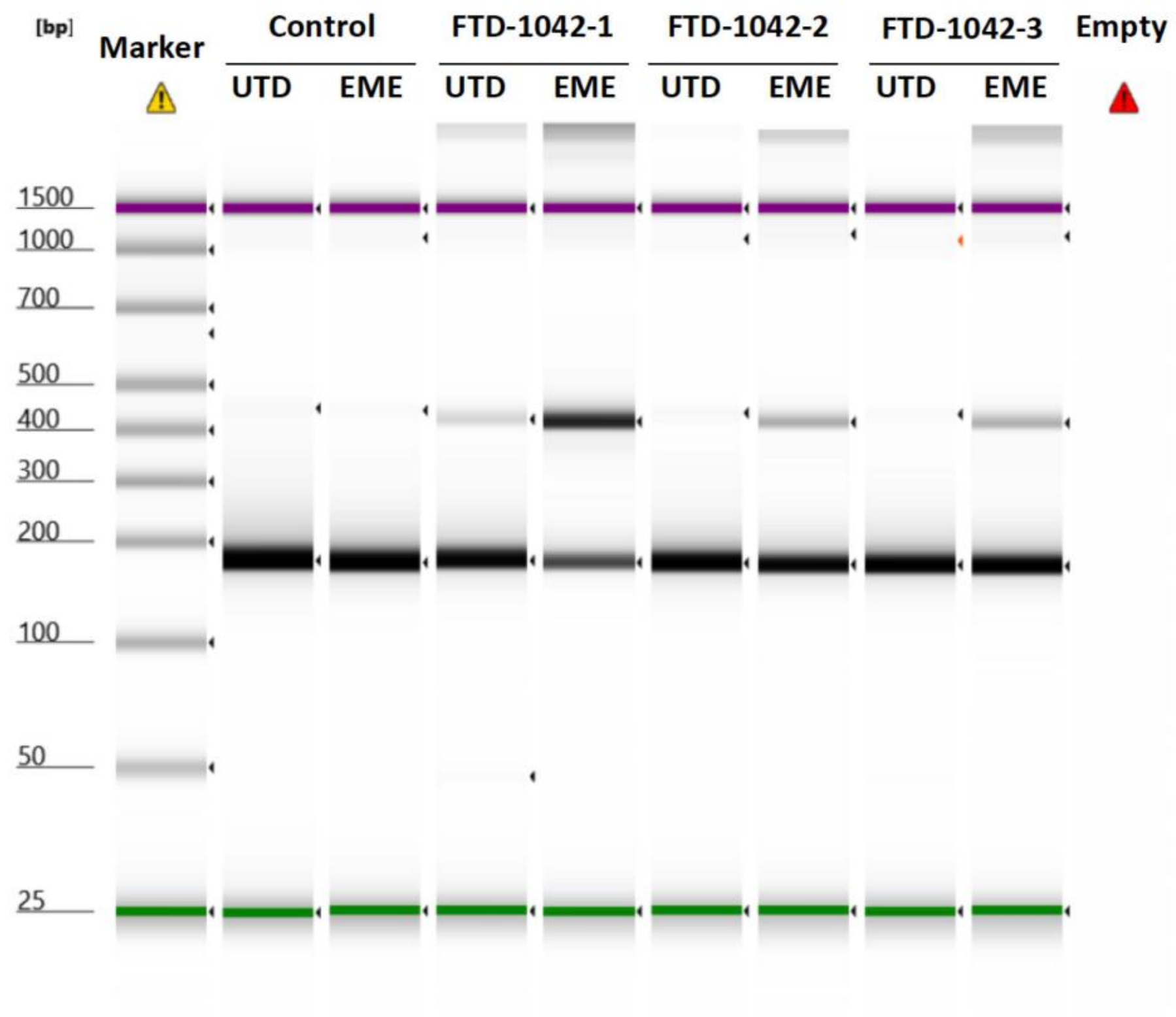

**Supplemental Figure S3. Full-length gel corresponding to the cropped gel on figure 4A.** LCL of patients from the family FTD-1042 were previously untreated (UTD) or treated with emetine (EME). FTD-1042-1 is homozygous for the *GRN* mutation c.709-3C>G. FTD-1042-2 and FTD-1042-3 are heterozygous.

| Name | Immunogen epitope | Monoclonal/polyclonal | Company |
|---|---|---|---|
| TDP-43 | Glu204, Asp205 and Arg208 of TDP 43 | Mouse monoclonal clone 6H6E12 | Proteintech |
| Phosphorylated TDP-43 | pS409/410 of TDP-43 | Mouse monoclonal | Cosmo Bio |
| Ubiquitin | Recombinant full length protein corresponding to human ubiquitin | Rabbit polyclonal | DAKO |
| Aβ | Residues 9-14 of Aβ | Mouse monoclonal 6F/3D | DAKO |
| Phosphorylated tau | pS202/pT205 & S208 of the tau molecule | Mouse monoclonal Clone AT8 | ThermoFisher |
| A-Synuclein | Residues 47-53 of α-synuclein | Mouse monoclonal clone 5G4 | Analytikjena |
| Double labeling Neurofilament | 70 kDa subunit of neurofilament | Mouse monoclonal Clone 2F11 | Dako |
| Myelin basic protein (MBP) | Recombinant protein part of MBP | Rabbit monoclonal | Diagomics |

**Supplementary table S1. Antibodies used for IHC studies on brain tissue.**

**Supplementary material**

**Plasma progranulin dosage**

Plasma progranulin levels were measured by ELISA method using progranulin-human-Elisa kit (Adipogen, Coger SAS, France), according to the manufacturer's instructions. The analytical performance of the kit was previously verified in our laboratory by comparing of progranulin levels in plasma from 85 FTD patients carrying heterozygous *GRN* mutations to 50 age matched controls (unpublished data). The optimal ROC cut-off values at 72 µg/L, as calculated by Youden index, discriminated between *GRN* -/+ and controls with 97.6% sensitivity and 100 % specificity (AUC = 0.998, 95% CI [0.995-1.003]). PPV = 100% and NPV = 96%. We used the test for diagnosis screening before molecular screening in clinical practice, and have set a cut-off value of 72 µg/L defining the limit between normal and abnormal levels of plasma progranulin with 96 % sensitivity and 100 % specificity. (the cutoff value maximizing the sensitivity was of 91 µg/ml (100 % sensitivity, 90 % specificity)).

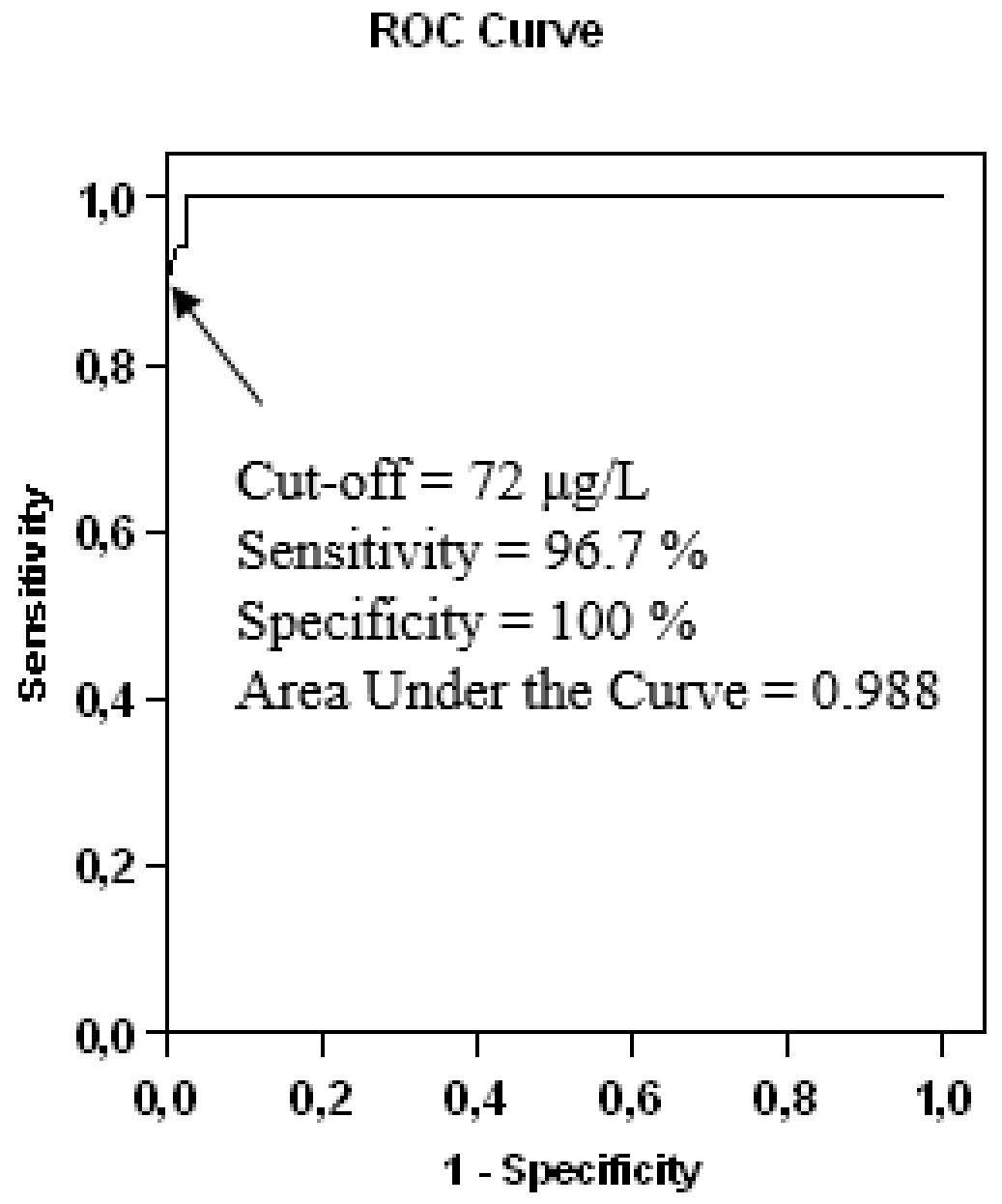